\documentclass[superscriptaddress,secnumarabic,
amssymb,amsmath,nobibnotes,aps,prd,showkeys,showpacs,nofootinbib]{revtex4}%
\usepackage{graphicx}
\usepackage{epsf}
\usepackage{bm}
\usepackage{amsmath}
\usepackage{amsfonts}
\usepackage{amssymb}
\usepackage{epstopdf}
\usepackage{subfigure}
\usepackage{natbib}
\usepackage{color}%
\usepackage{hyperref}
\hypersetup{dvips,dvipdfm,linktoc=page,colorlinks=true,linkcolor=blue,citecolor=red,filecolor=magenta,urlcolor=magenta,bookmarks=true}
\providecommand{\U}[1]{\protect\rule{.1in}{.1in}}
\newcommand{\be}{\begin{equation}}
	\newcommand{\ee}{\end{equation}}

\newcommand{\mincir}{\raise
	-3.truept\hbox{\rlap{\hbox{$\sim$}}\raise4.truept\hbox{$<$}\ }}
\newcommand{\magcir}{\raise
	-3.truept\hbox{\rlap{\hbox{$\sim$}}\raise4.truept\hbox{$>$}\ }}

\usepackage{tikz}

\begin{document}
	\title{ A study of interacting scalar field model from the perspective of the dynamical systems theory}

\author{Goutam Mandal}
\email{goutam.math.nbu@gmail.com ; rs\_goutamm@nbu.ac.in}
\affiliation{Department of Mathematics, University of North Bengal, Raja Rammohunpur, Darjeeling-734013, West Bengal, India.}
\author{Soumya Chakraborty}
\email{soumyachakraborty150@gmail.com}
\affiliation{Department of Mathematics, Jadavpur University, Kolkata-700 032, West Bengal, India}
\author{Sudip Mishra}
\email{sudipcmiiitmath@gmail.com}
\affiliation{Department of Applied Sciences, Maulana Abul Kalam Azad University of Technology, Haringhata, Nadia- 721249, West Bengal, India}
\author{Sujay Kr. Biswas}
\email{sujaymathju@gmail.com; sujay.math@nbu.ac.in}
\affiliation{Department of Mathematics, University of North Bengal, Raja Rammohunpur, Darjeeling-734013, West Bengal, India.}

\begin{abstract}
In this work, considering the background dynamics of flat Friedmann-Lemaitre-Robertson-Walker(FLRW) model of the universe, we investigate a scalar field model as dark energy candidate which interacting with the pressure-less dust as dark matter from dynamical systems perspective. From  phenomenological vantage point two interaction terms are chosen: one depends on Hubble parameter $H$ and other is local, independent of Hubble parameter. In interaction model 1, the scalar field potential as well as the coupling are considered to be in the form of inverse square and accordingly a two-dimensional autonomous system is obtained.
On the other hand, Interaction model 2 comprises with the potential as well as coupling of scalar field which are considered in form of exponential function of scalar field ($\phi$) and as a result of which a four-dimensional autonomous system is achieved. We study two systems separately and come by several critical points in 2D system as well as in 4D system.  We have derived sound speed and the classical stability conditions.  Furthermore, for 2D autonomous system we analyzed the stability of some critical points at infinity.
From this autonomous system, we obtain scalar field dominated solutions representing late time accelerated evolution of the universe that  does not elucidate the coincidence problem. Late time scaling solutions are also realized by the accelerated expansion of the universe which evolves in quintessence era that alleviates the coincidence problem successfully. From the analysis of 4D system, we obtain non-hyperbolic sets of critical points which are analyzed by the center manifold theory. In this model, the de Sitter like solutions represent the transient evolution of the universe.
\end{abstract}

\maketitle
\section{Introduction}
Theoretical cosmology does not agree with the recent observational evidences \cite{Shahalam:2015lra, Melnikov:2010zza} which predict that currently the Universe is enduring an accelerated expansion.  So, cosmologists have been trying to explain these observational predictions for more than one decade separated in two groups.  One group of cosmologists has been trying to explain the observational facts by introducing an exotic quantity commonly known as dark energy (DE) with large negative pressure, while the other group is focused to explain the observational evidences theoretically by modifying the Einstein gravity theory.\\

In the literature, there are various choices of DE but \textit{cosmological constant} is observationally more preferable as well as the simplest one.  But this choice also consists of two severe conceptual problems, 
(1) cosmological constant problem and (2) coincidence problem. On the other hand, there is no well-accepted modification of gravity theory as an alternative choice to the cosmologists. Cosmological constant problem can be overcome by introducing new type of time varying DE model in which the canonical scalar field model: \textit{quintessence} is the popular one where equation of state parameter $\omega_{\phi}$ for quintessence takes any value in the range $-1<\omega_{\phi}<-\frac{1}{3}$. After that various DE models based on scalar field such as K-essence, phantom, quintom etc. have been studied in the literature. The scalar field model, where energy density and pressure having a non-canonical form of kinetic term, i.e.,  kinetic term appears with a coupling function (non-canonical term) depending on scalar field $\phi$, is the non-canonical scalar field model \cite{Fang,Mukhanov,Mamon1,Mamon2,Dutta2019} which has also been studied extensively to solve several cosmological issues \cite{Lee}. On the other hand, coincidence problem can be alleviated by introducing an interaction term in between the dark energy and the dark matter. Since at present the evolution of the universe is assumed to be dominated by the dark energy and the dark matter, the possibility of having interaction between them cannot be ignored. Also, an appropriate form of interaction can provide a possible mechanism to alleviate the coincidence problem. However, due to unknown nature of DE, one can choose the interaction term phenomenologically which can be a function of energy density of either DE or DM or both \cite{Olivares}. Several interacting DE models where interaction terms are taken in phenomenological background have been investigated in the literature (see refs. \cite{Yuri.L.Bolotin2014,Andre A.Costa2014,M.Khurshudyan2015,N.Tamanini2015,T.Harko2013,Nunes2016,Wang2016,Copeland1,Aguirregabiria1,Lazkoz1,Fang1,Leon1,Leon2,Odintsov2018a,Odintsov2018b,Bahamonde2018}). \\

Various observational data suggests that the Universe is mainly dominated by a pressureless dark matter (DM) and dark energy (DE) where DM contributes around $28 \%$ and DE contributes around $68 \%$ of the total energy budget of the Universe and these form the dark sector \cite{Planck}. $\Lambda CDM$ model is assumed to be the standard cosmological model which is capable of explaining the late time acceleration of the Universe. But, it has two serious issues. One is the cosmological constant problem which arises due to the disagreement between the observed small value of the cosmological constant and the theoretical large value of vacuum suggested by the quantum field theory. The second one is cosmic coincidence problem. This is the basic motivation for Researchers to find some other cosmological models free from the two problems. Since, the nature of the components in the dark sector is not clear yet, an interaction in the dark sector components can be chosen in order to alleviate those problems. The concept of interaction mechanism came into focus when some models providing explanations to the cosmic coincidence problem \cite{Amendola,Tocchini,Amendola1,del,Campo} were found. Also, in literature one can find solutions to the cosmological constant problem \cite{Wetterich} which motivates scientific communities to propose an interaction term in cosmological models. Moreover, an interaction in the dark sector can alleviate $H_{0}$ tension \cite{Di,Kumar,Yang,Pan,Kumar1,E,Akarsu,Lucca,Valentino,L,Renzi,The,Vagnozzi2020,Visinelli2019,Valentino2020,Cheng2020,Vagnozzi2020a,Pan2020,Yang2021,Gao2021,Lucca2021,Nunes2021,Guo2021,Mancini2022,Ferlito2022,Nunes2022,Pan2022} which appears due to disagreement between the CMB measurements by Planck satellite within $\Lambda CDM$ cosmology \cite{Planck} and SH$0$ES \cite{Riess} and $S_{8}$ tension \cite{Pour,Rui,Kumar2,Araujo,Avila} which appears due to discordance (i.e. many sigmas gap) between the Planck and Weak Lensing measurements \cite{Valentino1} . In spite of all these, an interaction in the dark sector is capable of explaining the phantom phase of the DE without any need of a scalar field with negative correction \cite{Pan1,Bonilla,Bonilla1} . Therefore, based on these motivations, interaction mechanism in the dark sector achieves large attention in cosmological study in recent time. For comprehensive study on interacting dark energy models, we shall recommend two works \cite{Yuri.L.Bolotin2014,Wang2016} already in the literature.\\


Motivated from the above fact, in the present manuscript,   cosmological implications of a scalar field model have been studied in the background of homogeneous and isotropic flat FLRW space-time.  The system of cosmological equations are nonlinear differential equations.  There is no well known method to find the exact solution of the system.  Dynamical system analysis is an elegant tool to study nonlinear systems.  It allows us to gain a qualitative understanding of any cosmological model. The discussions on this topic are appeared in the previous references \cite{Yuri.L.Bolotin2014,Andre A.Costa2014,M.Khurshudyan2015,N.Tamanini2015,T.Harko2013,Nunes2016,Wang2016,Copeland1,Aguirregabiria1,Lazkoz1,Fang1,Leon1,Leon2,Odintsov2018a,Odintsov2018b,Bahamonde2018,Burin2005,Cabral2009} and also in a review book on cosmological dynamical system \cite{LeonBook2012}. 
Apparently many recent research works are going on this approach to understand cosmological models both geometrical and physical point of view.  To survey some of the current literatures we  cite papers \cite{Mishra:2019kfh, Mishra:2019uqm, Mishra:2019vnv, Mishra:2018fdx, Aljaf2020, Biswas:2020een,soumya} where authors studied cosmological evolutions of the Universe by dynamical system analysis.  By suitable change of variables Einstein field equations are transformed to an autonomous set of equations in terms of variables in a 2D and 4D phase space for various choices of interactions.   So the question arises that whether the cosmological solution of the scalar field model is stable when subject to perturbation in the phase space.   The fluctuation, near a hyperbolic critical point \cite{Biswas:2016idx,Biswas:2015cva}, can not change much about the stability but a small perturbation near a non-hyperbolic critical point appears to be different stability criteria. So it is important to study non-hyperbolic critical points.  As Hartman-Grobman theorem is not appropriate to characterize non-hyperbolic critical points,  we thoroughly study center manifold theory and apply it to characterize the non-hyperbolic critical points. We have derived sound speed and the classical stability conditions.  Furthermore, we find some extra fixed points for which the dynamical variables start to diverge to very high values.  This situation happens when the trajectory of the universe represents possible states in the asymptotic past in the finite phase space. We study the physical significance or novelty of the corresponding results.\\

From the scalar-tensor theory, and for mathematical simplicity we choose two interactions: one is depending directly on Hubble parameter $H$ and energy density of matter $\rho_{m}$, another is chosen locally, depending only on energy density $\rho_{m}$ and independent of $H$. For the first interaction model, the potential and the coupling function (non-canonical term) of the scalar field are considered as inverse square form. As a result, autonomous system of dynamical system is reduced to a two dimensional. On the other hand, for interaction model 2, the potential and the coupling function are considered to be depended exponentially on scalar field $\phi$. This leads to a four dimensional autonomous system. In 2D system, for interaction model 1, we obtain some interesting critical points including hyperbolic type and normally hyperbolic set of critical points. Linear stability theory shows that accelerated scalar field dominated solutions are the late time attractor. This study also reveals that the accelerated scaling solutions are attractors in quintessence era. Further, a detailed center manifold theory is employed in the interaction model 2 which exhibits the non-hyperbolic sets of critical points being the saddle-like solutions in phase space provide the transient nature of evolution whenever they are scalar field dominated or scaling like solutions in the phase space.\\

We organize our work in the following manner: In the next section \ref{formation_autonomous}, we start with scalar field model, by choosing phase space variables we construct the autonomous system of ordinary differential equations. Phase space analysis of the autonomous system by considering two interaction models with different potential and coupling have been presented in the section \ref{phase_space_auto}. After that center manifold theorem has been studied for non-hyperbolic type critical points in \ref{CMT}. We expound compactification and dynamics around the points at infinity in section \ref{infinity}. We present cosmological implications in a section \ref{Cosmological Implications}. Finally, We conclude with a brief discussion in \ref{conclusions}.



\section{ Basic equations of scalar field and construction of autonomous system}
\label{formation_autonomous}
Here, we consider a model where gravity is minimally coupled to scalar field but the scalar field is not canonical. There is a pre-factor in the kinetic term of the scalar field. So, the scalar field Lagrangian \cite{Fang,Mukhanov,Mamon1,Mamon2,Dutta2019,Bertacca:2007,Bertacca:2008,Bose:2009,Chimento:2004,Santiago:2011,Mishra:2019kfh} of this system refers to the following action

\begin{equation}
	A= \int d^4x\sqrt{-g}\bigg[\frac{R}{2\kappa^2}-\frac{1}{2}\lambda(\phi)g^{\mu\nu}(\nabla_\mu\phi)(\nabla_\nu\phi)-V(\phi)\bigg]+A_m, \label{2.1}
\end{equation}
where  $A_m$ is the action corresponding to the matter content in the universe \cite{Mahata:2013oza}.  In the present case the matter is chosen as cold dark matter (CDM) in the form of dust.  $\lambda(\phi)$, the coupling function is chosen as arbitrary function of the scalar field $\phi$, $V(\phi)$ is the potential of the scalar field and $\kappa^2 = 8\pi G$ is the Gravitational Coupling parameter.  \\

The present cosmological model is considered in the background of homogeneous and isotropic flat Friedmann-Lema\^{i}tre-Robertson-Walker (FLRW) space-time manifold having line element
\begin{equation}
	ds^2=-dt^2+a^2(t)\left[dr^2+r^2\left(d\theta^2+sin^2\theta d\phi^2\right)\right].
\end{equation}
Friedmann equation and acceleration equation for scalar field in presence of pressureless dust in the background of FLRW metric (using natural units $\left(\kappa^{2}=8\pi G=1  \right)$) are:
\begin{equation}\label{Friedmann}
	H^{2} =\frac{1}{3}(\rho_{m}+\rho_{\phi})
\end{equation}
and
\begin{equation}\label{Raychaudhuri}
	2 \dot{H} = -(\rho_{m}+\rho_{\phi}+p_{\phi}),
\end{equation}
where $H=\frac{\dot{a}}{a}$ is Hubble parameter and $\rho_{m}$ is the energy density for dark matter (DM) taken as pressureless ($p_m=0$) dust with equation of state parameter $\omega_{m}=0$.
Non-canonical scalar field governed by its energy density $\rho_{\phi}$ and pressure $p_{\phi}$ which are combined with potential and kinetic terms as follows: 
\begin{equation}\label{energy density scalar}
	\rho_{\phi}=\frac{1}{2}\lambda(\phi)\dot{\phi}^2+V(\phi)
\end{equation}\\
and 
\begin{equation}\label{pressure scalar}
	p_{\phi}=\frac{1}{2}\lambda(\phi)\dot{\phi}^2-V(\phi),
\end{equation}
where $V(\phi)$ is the potential and $\lambda(\phi)$ describes the non-negative, smooth coupling function of scalar field $\phi$. This scalar field reduces to canonical quintessence scalar field. 
The equation of state parameter for scalar field is $\omega_{\phi}= \frac{p_{\phi}}{\rho_{\phi}}$. Total (effective) energy density and pressure for the model 
$\rho_{eff}=\rho_{m}+\rho_{\phi}$ and $p_{eff}=p_{\phi}$ constitute the total (effective) equation of state parameter in the form:
\begin{equation}\label{equations}
	\omega_{eff}=\frac{p_{eff}}{\rho_{eff}}= \frac{p_{\phi}}{\rho_{m}+\rho_{\phi}}.
\end{equation}
In an interacting scenario when DE interacts with DM through an interaction term $Q$, the individual component (DE and DM) conserves separately and hence the conservation equations for scalar field and matter (with the term $Q$) take the following form :
\begin{equation}\label{continuity-scalar}
	\dot{\rho}_{\phi}+3H(\rho_{\phi}+p_{\phi})=-Q
\end{equation}
and
\begin{equation}\label{continuity-matter}
	\dot{\rho}_{m}+3H\rho_{m}=Q,
\end{equation}
where the sign of interaction term $Q$ specifies the direction of energy flow in the dark sector.
For example, $Q>0$ indicates the flow of energy transfers from DE to DM while for $Q<0$ the flow of energy transfers from DM to DE.
Now, evolution equation for the scalar field (from Eq. (\ref{continuity-scalar}), using Eq. (\ref{energy density scalar}) and Eq. (\ref{pressure scalar})) with coupling term takes the form 
\begin{equation}\label{EvolutionEqn-scalar}
	\frac{1}{2}\lambda'\dot{\phi}^2+\lambda\ddot{\phi}+V'+3H\lambda\dot{\phi}=-\frac{Q}{\dot{\phi}},
\end{equation}  
where the symbol `prime' denotes the derivative with respect to scalar field ($'\equiv \frac{d}{d\phi}$) . 
Since the evolution equations are very much complicated in the interacting scenario it is quite impossible to obtain an exact analytical solution from the Friedmann equation, acceleration equation and conservation. We try to obtain a qualitative picture of this cosmological model, we adopt dynamical system approach and for that choose the following dynamical variables 
\begin{equation}\label{dynamical_variables}
	x=\frac{\dot{\phi}}{\sqrt{2}},~~~y=\frac{\sqrt{V(\phi)}}{\sqrt{3}H},~~~z=\frac{\sqrt{\lambda(\phi)}}{\sqrt{3}H}.
\end{equation}  
For convenience, we have chosen the above auxiliary variables so that the kinetic energy and the potential energy become independent variables throughout the dynamical system analysis. 
One may note that the choice of dynamical variables is not unique.  However, if we suitably choose another set of auxiliary variables, the autonomous system will be in different form which may result distinct phase space dynamics but the overall physical interpretation remains unaltered. One may note that if any orbit connects $H<0$ with $H>0$, then our choice of variables put up with singularity. In this case, one may normalize the dimensionless variables not with $H$ but with $D$, where the expression of $D$ can be found in Ref. \cite{Leon2015}.

Using the variables (in equation (\ref{dynamical_variables})) the evolution equations reduce (after some algebra) to the following system of ordinary differential equations:
\begin{eqnarray}
	\begin{split}
		\frac{dx}{dN}& = -3x-\sqrt{\frac{3}{2}}\frac{\lambda'}{\lambda\sqrt{\lambda}} x^2z-\sqrt{\frac{3}{2}}\frac{V'}{V\sqrt{V}}\frac{y^3}{z^2}-\frac{Q}{6xz^2H^3},& \\
		\frac{dy}{dN}& = y\left[\sqrt{\frac{3}{2}}\frac{V'}{V\sqrt{V}}xy+\frac{3}{2}\left(1+x^2z^2-y^2\right)\right],& \\
		\frac{dz}{dN} & = z\left[\sqrt{\frac{3}{2}}\frac{\lambda'}{{\lambda}\sqrt{\lambda}}xz+\frac{3}{2}\left(1+x^2z^2-y^2 \right)\right].&
		~~\label{Gen_autonomous}
	\end{split}
\end{eqnarray}
Here, $ N=\ln a $ is chosen as the independent variable. Now, the system (\ref{Gen_autonomous}) cannot be a complete autonomous dynamical system of ODEs for the dynamical variables $x(N)$, $y(N)$ and $z(N)$ due to involvement of the terms containing  $\frac{V'}{V\sqrt{V}}$ and $\frac{\lambda'}{{\lambda}\sqrt{\lambda}}$ which are explicit functions of scalar field $\phi$. However, to accomplish dynamical system analysis we convert the terms into dynamical variables (by setting $s(\phi)=\frac{V'}{V\sqrt{V}}$ and $u(\phi)=\frac{\lambda'}{{\lambda}\sqrt{\lambda}}$). As a result, the time evolution equations of $s(\phi)=\frac{V'}{V\sqrt{V}}$ and $u(\phi)=\frac{\lambda'}{{\lambda}\sqrt{\lambda}}$ are as follows:
\begin{equation}\label{s_equation}
	\frac{ds}{dN}=\sqrt{6}xys^2 \left(\varGamma_s -\frac{3}{2}\right),
\end{equation} 

\begin{equation}\label{u_equation}
	\frac{du}{dN}=\sqrt{6}xzu^2 \left(\varGamma_u -\frac{3}{2}\right)	,
\end{equation}
where $\varGamma_s=\frac{VV''}{V'^2}$ and $\varGamma_u=\frac{\lambda \lambda''}{\lambda'^2}$. Note that the system (\ref{Gen_autonomous}) along with the equations (\ref{s_equation}) and (\ref{u_equation}) forms a complete five dimensional autonomous system of ODEs for the dynamical variables $x(N)$, $y(N)$, $z(N)$, $s(N)$ and $u(N)$.  Fadragas {\it et al.} in the Ref. \cite{Fadragas2014} performed a comprehensive dynamical system analysis of anistropic scalar field cosmologies for a wide range of potentials. But in our model, 
	it is very much complicated to handle 5D autonomous system rather than 3D system and for inverse square potential $V\propto \phi^{-2}$, the terms $s^2 (\varGamma_s-\frac{3}{2})$ vanishes which implies $\frac{ds}{dN}=0$, i.e., when $s$=constant. Similarly, for coupling function $\lambda(\phi)\propto \phi^{-2}$, we get $u$=constant which implies that $\frac{du}{dN}=0$.  However, the potential diverges at $\phi=0$. It is to be noted that the potential to have a maximum at $\phi\longrightarrow 0$ and to die off for a large scalar field \cite{Feinstein}. This potential is able to produce a power-law expansion in classical cosmology. Here, Hubble parameter can be used instead of scalar field $\phi$ as a fundamental quantity by employing the Hamilton-Jacobi formulation \cite{Huang}.

 Thus, the resulting autonomous system will be of three dimensional system when $s$ and $u$ are non-zero constant parameters (for inverse square potential and coupling function) which we shall discuss in the next section.\\

Now, the cosmological parameters can be written in terms of dynamical variables as follows:\\
Density parameter for scalar field (DE) is
\begin{equation}\label{density_parameter}
	\Omega_{\phi}=x^2z^2+y^2,
\end{equation}
the equation of state parameter for scalar field (DE) is
\begin{equation}\label{eqn_of_state_parameter}
	\omega_{\phi}=\frac{x^2z^2-y^2}{x^2z^2+y^2},
\end{equation}
the effective equation of state parameter for the model is
\begin{equation}\label{eff_eqn_of_state_parameter}
	\omega_{eff}=x^2z^2-y^2,
\end{equation}
and the deceleration parameter is
\begin{equation}\label{dec_parameter}
	q=-1+\frac{3}{2}(1+\omega_{eff})
\end{equation}
One can obtain the conditions for acceleration by setting either $q<0$, or $\omega_{eff}<-\frac{1}{3}$
and for deceleration :~~~~~~$q>0$~~~~~i.e.~~~~~ $\omega_{eff}>-\frac{1}{3}.$
Friedmann equation (\ref{Friedmann}) gives the constraint equation in terms of density parameter for DM is
\begin{equation}\label{density_dm}
	\Omega_{m}=1-x^{2}z^2-y^2
\end{equation}
by which due to the energy condition $ 0\leq \Omega_{m} \leq 1 $, we obtain the constraints for dynamical variables in the region:
\begin{equation}
	0\leq x^{2}z^2+y^2 \leq 1\label{eqn18}.
\end{equation} 
For mathematical simplicity, we debar from the case where the evolution of the universe is kinetic dominated or scalar field dominated and that indicates that $x$ is confined in a domain of finite region.  We also preclude the possibility of scalar field $\phi$ locally or globally to be approximated by a constant function with respect to time `$t$' and that restricts $z$ in a domain of finite region from the inequality (\ref{eqn18}). \\

In the following section, we shall explicitly analyze the system (\ref{Gen_autonomous}) for two different choices of the interaction terms, namely,
$(i)~Q_1 \propto H\rho_{m}$~~and~~$(ii)~Q_2 \propto \rho_{m}$ with the potential $V(\phi)$ and the coupling function $\lambda (\phi)$ of the scalar field are considered as inverse square and exponential form respectively.\\



\section{Phase space analysis of the system (\ref{Gen_autonomous}) with different choices of interaction terms and potentials}
\label{phase_space_auto}

We shall now present a detailed phase space analysis of the general system (\ref{Gen_autonomous}), where first analyze the interaction $Q_1 \propto H\rho_{m}$  \cite{Chen,Bohmer Interation2a} with inverse square potential and coupling and then we analyze  $Q_2 \propto \rho_{m}$ \cite{Bohmer Interation2a,Bohmer Interaction2b} with the exponential form of potential and coupling function of the scalar field. Hyperbolic type critical points are analyzed by using the Hartman -Grobman theorem (linear stability theory) and the center manifold theory is applied for checking the stability of non-hyperbolic critical points.

\subsection{Interaction model 1: $Q_1: Q\propto H \rho_{m}$}

First, we consider the interaction term
\begin{equation}\label{Interaction1}
	Q=6\eta H\rho_{m},
\end{equation}
where $\eta$ is the coupling parameter measures the strength of the interaction. Positivity of the coupling parameter indicates the energy transfers from DE to DM which is required to alleviate the coincidence problem and is compatible with second law of thermodynamics.
Using the term of (\ref{Interaction1}) in the system (\ref{Gen_autonomous}) with inverse square form of potential and coupling function of scalar field, i.e., 
$V(\phi)=V_{0}\phi^{-2}~~\mbox{and}~~\lambda(\phi)=\lambda_{0}\phi^{-2}$ 
such that $s$ and $u$ become constant parameters which leading  to a 3D autonomous system as $\frac{ds}{dN}=0$ in equation (\ref{s_equation}) and $\frac{du}{dN}=0$ in equation (\ref{u_equation}). 
Note that for $V_0=\lambda_{0}$ the autonomous system leads to a two dimensional system because the dynamical behavior of two variables $y$ and $z$ in eq.(\ref{dynamical_variables}) are same. As a result, in the study of this model with inverse square potential term for both the potential ($V(\phi)$) and coupling ($\lambda(\phi)$), we shall study only 2D dynamical system by replacing the variable $z$ with $y$, and the constant parameter $u$ with $s$. 
Now, 2D autonomous system in $x-y$ takes the following form:
\begin{eqnarray}
	\begin{split}
		\frac{dx}{dN}& =-3x-\sqrt{\dfrac{3}{2}}sx^2y-\sqrt{\dfrac{3}{2}}sy-\frac{3\eta}{xy^2}(1-x^2y^2-y^2),& \\
		\frac{dy}{dN}& =y\left[\sqrt{\frac{3}{2}}sxy+\frac{3}{2}\left( 1+x^2y^2-y^2\right )  \right]. &~\label{auto-Int-1}
	\end{split}
\end{eqnarray}
The system (\ref{auto-Int-1}) has singularities at $x=0$~~and~~$y=0$ respectively. To remove these, the right hand side (r.h.s) of each equation of the system (\ref{auto-Int-1}) is multiplied by $ x^2y^2 $. Then the system (\ref{auto-Int-1}) reduces to the form :~~\\
\begin{eqnarray}
	\begin{split}
		\frac{dx}{dN}& =-3x^3y^2-\sqrt{\frac{3}{2}} s x^4 y^3-\sqrt{\frac{3}{2}}sx^2y^3-3\eta x(1-x^2y^2-y^2) ,& \\
		\frac{dy}{dN}& =x^2y^3 \left[ \sqrt{\frac{3}{2}}sx y+\frac{3}{2} \left( 1+x^2 y^2-y^2\right) \right] ,&~\label{autonomous2D}
	\end{split}
\end{eqnarray}

The critical points extracted from the system (\ref{autonomous2D}) are the following:
{\bf 
	\begin{itemize}
		\item  I.  Critical point : $ A_{1}=(0,y_{c})$
		\item  II. Critical point : $B_{1}=\left(-\frac{s}{\sqrt{6-s^2}},~~ \sqrt{1-\frac{s^2}{6}}\right) $
		\item  III. Critical point : $ C_{1}=\left(\frac{s}{\sqrt{6-s^2}},~~ -\sqrt{1-\frac{s^2}{6}}\right) $
		\item  IV. Critical point : $ D_{1}=\left(\frac{\sqrt3 (2\eta-1)}{\sqrt{4\eta s^2 +3(2\eta-1)^2}},~~\frac{\sqrt{4\eta s^2 +3(2\eta-1)^2}}{\sqrt{2} s}\right)$
		\item  V. Critical point : $ E_{1}=\left(-\frac{\sqrt3 (2\eta-1)}{\sqrt{4\eta s^2 +3(2\eta-1)^2}},~~-\frac{\sqrt{4\eta s^2 +3(2\eta-1)^2}}{\sqrt{2} s} \right)$
	\end{itemize}
	
}


\subsubsection{Phase space analysis of interaction 1}
\begin{figure}
	\centering
	\subfigure[]{%
		\includegraphics[width=7cm,height=6cm]{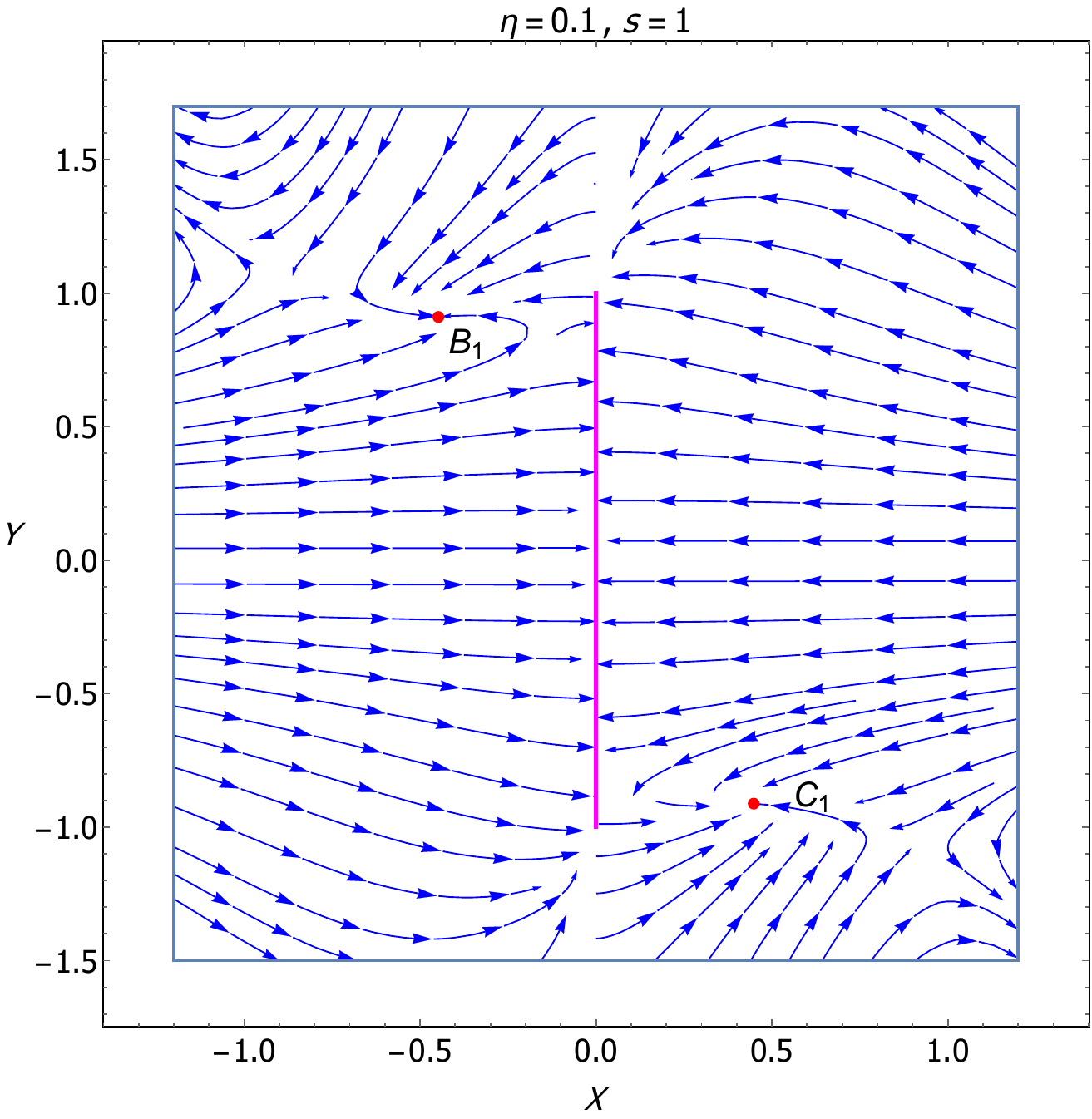}\label{A1stable}}
	\qquad
	\subfigure[]{%
		\includegraphics[width=7cm,height=6cm]{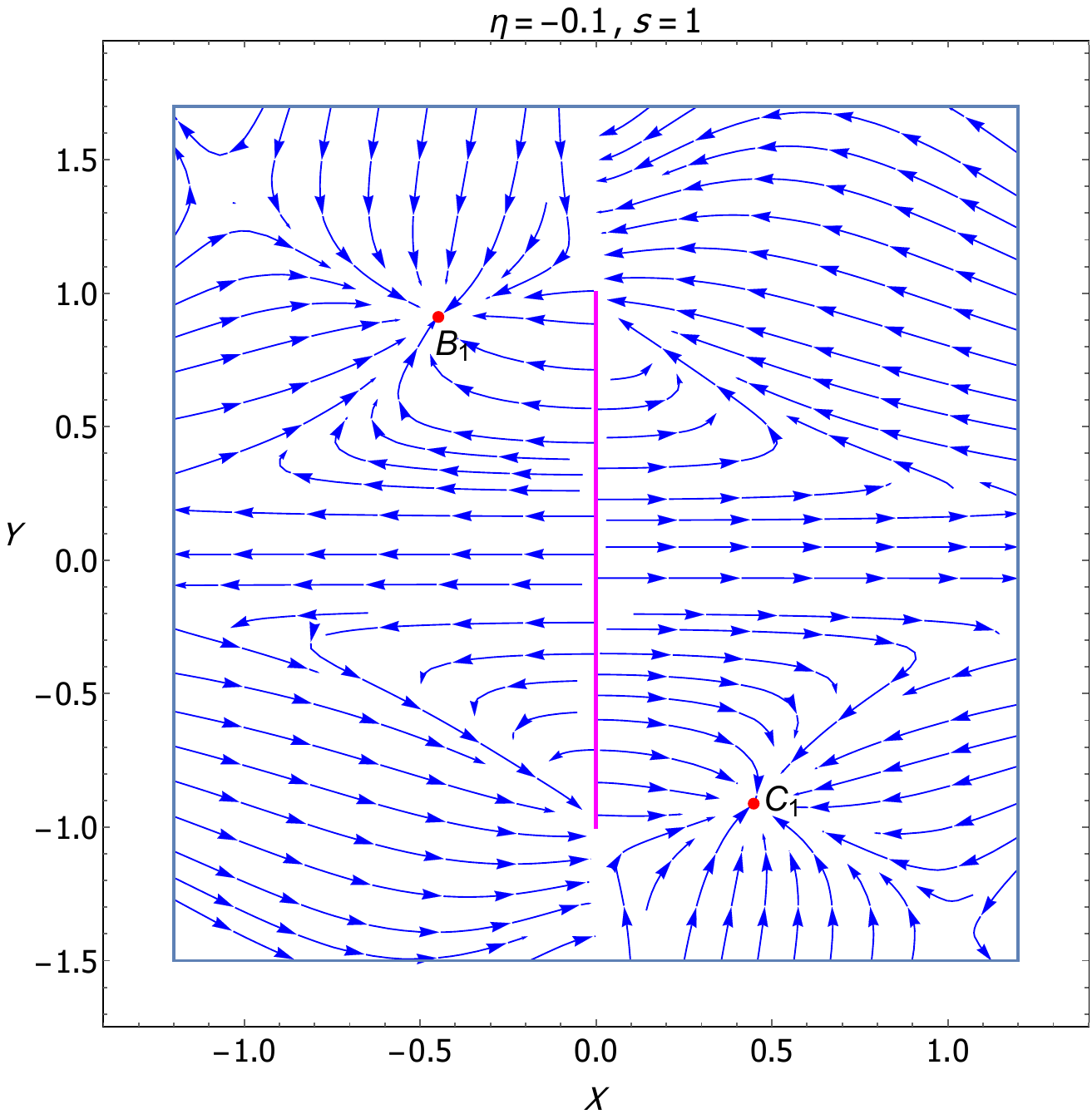}\label{A1unstable}}
	\qquad
	\subfigure[]{%
		\includegraphics[width=10cm,height=6cm]{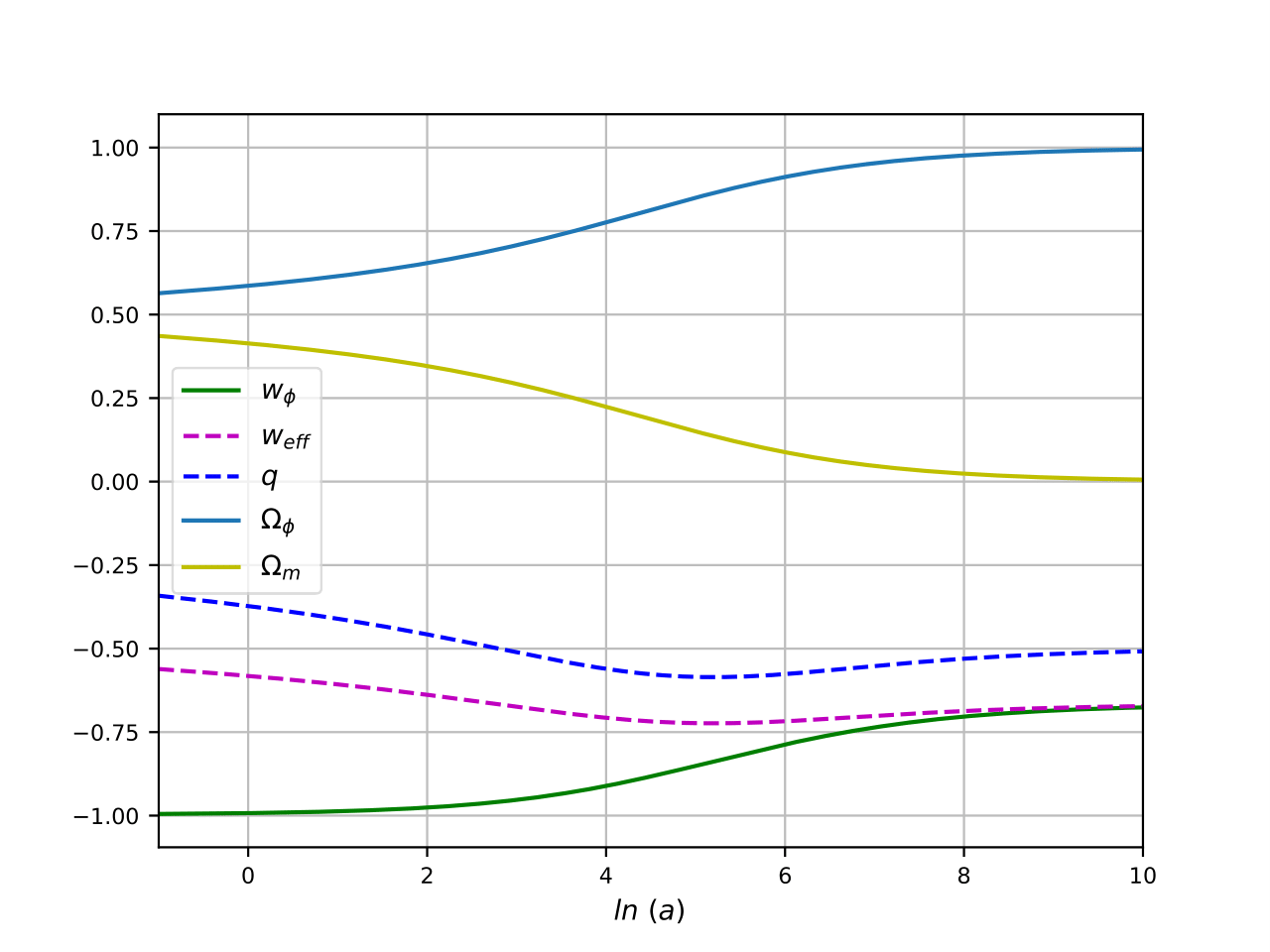}\label{Evolution_C}}
	\caption{The figure shows stability of critical points in $x-y$ plane and time evolution of cosmological parameters for the Interaction model 1. In panel (a) vector field shows that late time stable solutions $A_1$, $B_1$ and $C_1$ are attracted in quintessence era for $\eta=0.1$ and $s=1$ where the set $A_1$ magenta line represented by the magenta line. In panel (b), for parameter values $\eta=-0.1$ and $s=1$ the point $B_1$ and $C_1$ are stable and the set $A_1$ magenta colored is unstable. Finally, evolution of cosmological parameters for are in panel (c) for $\eta=0.01$ and $s=-1$ showing that the late time accelerated solutions attracted in quintessence era.}
	\label{phasespace-figure-CD}
\end{figure}

The autonomous system of ODEs (\ref{autonomous2D}) extracts five critical points, one is set of points and the remaining are isolated critical point. Phase plane analysis of these critical points are given below. Critical points and their corresponding physical parameters are shown in the table (\ref{Physical parameters model1}).
\begin{itemize}
	\item 
	Set of critical points $A_1$ exists for all parameter values $\eta$ and $s$ on the $y$-axis satisfying $0\leq y^{2}_c\leq 1$. The expressions of physical parameters corresponding to the set of points are shown in Table \ref{Physical parameters model1} suggesting that the set corresponds to solution having both of the DE and DM components in it. The DE associated with this set behaves as cosmological constant like fluid. Each point on this set exhibiting the accelerated universe when $y^{2}_c>\frac{1}{3} $. The Eigenvalues of first order perturbed matrix for the set $A_1$ are: $\{\mu_1=0~~\mbox{and}~~ \mu_2=3 \eta\left(  y_{c}^2 -1\right) \}$. Therefore, the points on this set are non hyperbolic critical points. It is to be noted that the set $A_1$ is normally-hyperbolic in nature because it has exactly one vanishing eigenvalue. 
	The stability of normally-hyperbolic set can be completely classified by considering the sign of the non-zero eigenvalues in the corresponding eigen-directions.  This statement can also be verified by center manifold theory as we will show it in Eq.(\ref{flow A1}). The stability depends on the flow parallel to the  eigen-direction of $\mu_1$ (i.e. x-axis) which is shown in the next section \ref{CMT}.
	The set can behave as completely DM dominated for $y_c \longrightarrow 0$ while the set represents the DE dominated (completely) solution as $y_c \longrightarrow 1$. Interestingly, for $y_c=1$, the point $(0,1)$ on the set $A_1$ describes the de Sitter expansion of the universe [$\Omega_{m}=0, \Omega_{\phi}=1, \omega_{eff}=q=-1$], but for this case the eigenvalue $\mu_2$ becomes zero and consequently dynamics near the point becomes very much complicated. However, the points always evolve in quintessence era. On the other hand, for $y_c=0$, the set will become completely DM dominated solutions but stability shows the set enables its stability when $\eta>0$ and unstability for $\eta<0$ respectively. Physical parameters for this case are : $[\Omega_{m}=1, \Omega_{\phi}=0, \omega_{eff}=q=0]$. Figure \ref{A1stable} shows the set $A_1$ is stable for $\eta>0$ and figure \ref{A1unstable} shows $A_1$ is unstable for $\eta<0$.

	\item
	Critical points $B_1$ and $C_1$ exist for $-\sqrt{6}<s<\sqrt{6}$ and for all values of coupling parameter $\eta$. The critical points are completely dominated by scalar field which behaves as any perfect fluid model (since $\omega_{\phi}=\frac{s^2}{3}-1$). Therefore, DE can represent either quintessence or cosmological constant or any other exotic type fluid depending on parameter $s$. Specifically, the restriction $s^2<2$ ($s\neq 0$) implies that the scalar field behaves as quintessence like fluid and there exists an accelerating universe (since for this case  $\omega_{eff}<-\frac{1}{3},~ q<0$) near these critical points. Further, for $s=0$, scalar field behaves as cosmological constant and for that case evolution of the points characterise the de Sitter expansion ($\omega_{\phi}=-1,~ \omega_{eff}=-1,~ q=-1$) of the universe. Eigenvalues of the linearized Jacobian matrix for the points $B_1$ and $C_1$ are:
	$\left\lbrace  \mu_1=\frac{1}{12} s^2 \left(s^2-6\right)~~\mbox{and}~~\mu_2=\frac{1}{6} s^2 \left(6 \eta +s^2-3\right)\right\rbrace $.
	Critical points $B_1$ and $C_1$ are hyperbolic in nature and these are stable attractor for
	
	(i)~$\eta \leq -\frac{1}{2}~~\mbox{and}~~ \left(-\sqrt{6}<s<0~~\mbox{or}~~ 0<s<\sqrt{6}\right)$, or\\
	(ii)~$ \left(-\frac{1}{2}<\eta <\frac{1}{2}~~\mbox{and}~~ \left(-\sqrt{3-6 \eta }<s<0~~\mbox{or}~~ 0<s<\sqrt{3-6 \eta }\right)\right)$.\\ 
	The critical points $B_1$ and $C_1$ evolve in quintessence era for the following restrictions:\\ 
	(i)~$\eta \leq \frac{1}{6}~~\mbox{and}~~ \left(-\sqrt{2}<s<0~~\mbox{or}~~ 0<s<\sqrt{2}\right)$, or\\
	(ii)~$\frac{1}{6}<\eta <\frac{1}{2}~~\mbox{and}~~ \left(-\sqrt{3-6 \eta }<s<0~~\mbox{or}~~ 0<s<\sqrt{3-6 \eta }\right).$\\
	Therefore, one can conclude that the points $B_1$,$C_1$ correspond to a scalar field dominated accelerated attractor at late times.
	
	\item
	Scaling solutions represented by the critical points $D_1$ and $E_1$ exist in phase plane either for \\
	(i)~$-\frac{1}{2}<\eta <0~~\mbox{and}~~ \left(-\frac{1}{2} \sqrt{3} \sqrt{\frac{4 \eta-4 \eta ^2 -1}{\eta }}<s\leq -\sqrt{3-6 \eta }~~\mbox{or}~~ \sqrt{3-6 \eta }\leq s<\frac{1}{2} \sqrt{3} \sqrt{\frac{4 \eta-4 \eta ^2- 1}{\eta }}\right) $,~or\\
	(ii)~$ 0\leq \eta <\frac{1}{2}~~\mbox{and}~~ \left(s\leq -\sqrt{3-6 \eta }~~\mbox{or}~~ s\geq \sqrt{3-6 \eta }\right) $, ~or\\
	(iii)~$ \eta =\frac{1}{2}~~\mbox{and}~~ (s<0~~\mbox{or}~~ s>0).$\\
	
	Both the critical points have the ratio of the energy densities of DE and DM is $\frac{\Omega_{\phi}}{\Omega_{m}}=\frac{2\eta s^2 +3(2\eta-1)^2}{(2\eta-1)(3-s^2 -6\eta)}$. The scalar field behaves as any perfect fluid (as $\omega_{\phi}=-\frac{2\eta s^2}{2\eta s^2 +3(2\eta-1)^2}$). There exists acceleration for both the points if $\eta>\frac{1}{6}$. 
	For $\eta=\frac{1}{2}$ irrespective of $s$, the scalar field DE behaves as cosmological constant, and accelerated de Sitter solution exists for these critical points (since for this case: $\omega_{eff}=q=-1, \Omega_{\phi}=1, \Omega_{m}=0$). On the other hand, for uncoupled case $\eta=0$, scalar field DE behaves as dust ($\omega_{\phi}$=0) and dust dominated decelerating scaling solutions obtained ($\omega_{eff}=0, q=\frac{1}{2}, \Omega_{m}=1-\frac{3}{s^2}, \Omega_{\phi}=\frac{3}{s^2}$). \\
	
	Eigenvalues of the linearized Jacobian matrix for the critical points $D_1$,$E_1$:
	
	\begin{align*}
		\mu_1=-\frac{3 (2 \eta -1) \left(s^3 \left(4 \eta  \left(3 \eta +s^2-3\right)+3\right)^2 \left(4 \eta  \left(9 \eta +s^2-3\right)-3\right)+\Delta_{DE}\right) }{8 s^5 \left(4 \eta  \left(3 \eta +s^2-3\right)+3\right)^2}
		\\
		\mu_2=-\frac{3 (2 \eta -1) \left(s^3 \left(4 \eta  \left(3 \eta +s^2-3\right)+3\right)^2 \left(4 \eta  \left(9 \eta +s^2-3\right)-3\right)-\Delta_{DE}\right) }{8 s^5 \left(4 \eta  \left(3 \eta +s^2-3\right)+3\right)^2}
	\end{align*}
	\\
	where,\\
	\begin{align*}
		\Delta_{DE}= [~\left(-216 (2 \eta -1)^5+16 \eta ^2 s^6-24 \eta  (4 (\eta -3) \eta +5) s^4 -9 (1-2 \eta )^2 \left(60 \eta ^2-76 \eta +7\right) s^2\right)\times\\     
		\times s^4 \left(4 \eta  \left(3 \eta +s^2-3\right)+3\right)^4~]^{\frac{1}{2}}.
	\end{align*}
	The points $D_1$, $E_1$ represents stable attractor for:\\
	(i)~$-\frac{1}{2}<\eta <0$~~and~~ $\left(-\frac{1}{2} \sqrt{3} \sqrt{\frac{4 \eta-4 \eta ^2-1}{\eta }}<s<-\sqrt{3-6 \eta }~~\mbox{or}~~ \sqrt{3-6 \eta }<s<\frac{1}{2} \sqrt{3} \sqrt{\frac{4 \eta-4 \eta ^2-1}{\eta }}\right) $,~or\\
	(ii)~$ \eta =0~~\mbox{and}~~ \left(s<-\sqrt{3}~~\mbox{or}~~ s>\sqrt{3}\right)$, ~or\\
	(iii)~$0<\eta <\frac{1}{2}~~\mbox{and}~~ \left(-\frac{1}{2} \sqrt{3} \sqrt{\frac{4 \eta-12 \eta ^2+1}{\eta }}<s<-\sqrt{3-6 \eta }~~\mbox{or}~~ \sqrt{3-6 \eta }<s<\frac{1}{2} \sqrt{3} \sqrt{\frac{4 \eta-12 \eta ^2+1}{\eta }}\right).$\\
	Therefore, in the interacting scenario, the points $D_1$,$E_1$ represents the scaling attractor at late times when DE behaves as any perfect fluid, while for uncoupled case ($\eta=0$), DE behaves as dust and the points $D_1$, $E_1$ represents dust dominated decelerated stable solution (for $\eta=0, s^2>3$). So, uncoupled model can predict the future decelerated scaling solution by the points $D_1$, $E_1$ which is shown in Figure (\ref{FstableDeceleration}).

\end{itemize}	


\begin{figure}
	\centering
	\subfigure[]{%
		\includegraphics[width=7cm,height=6cm]{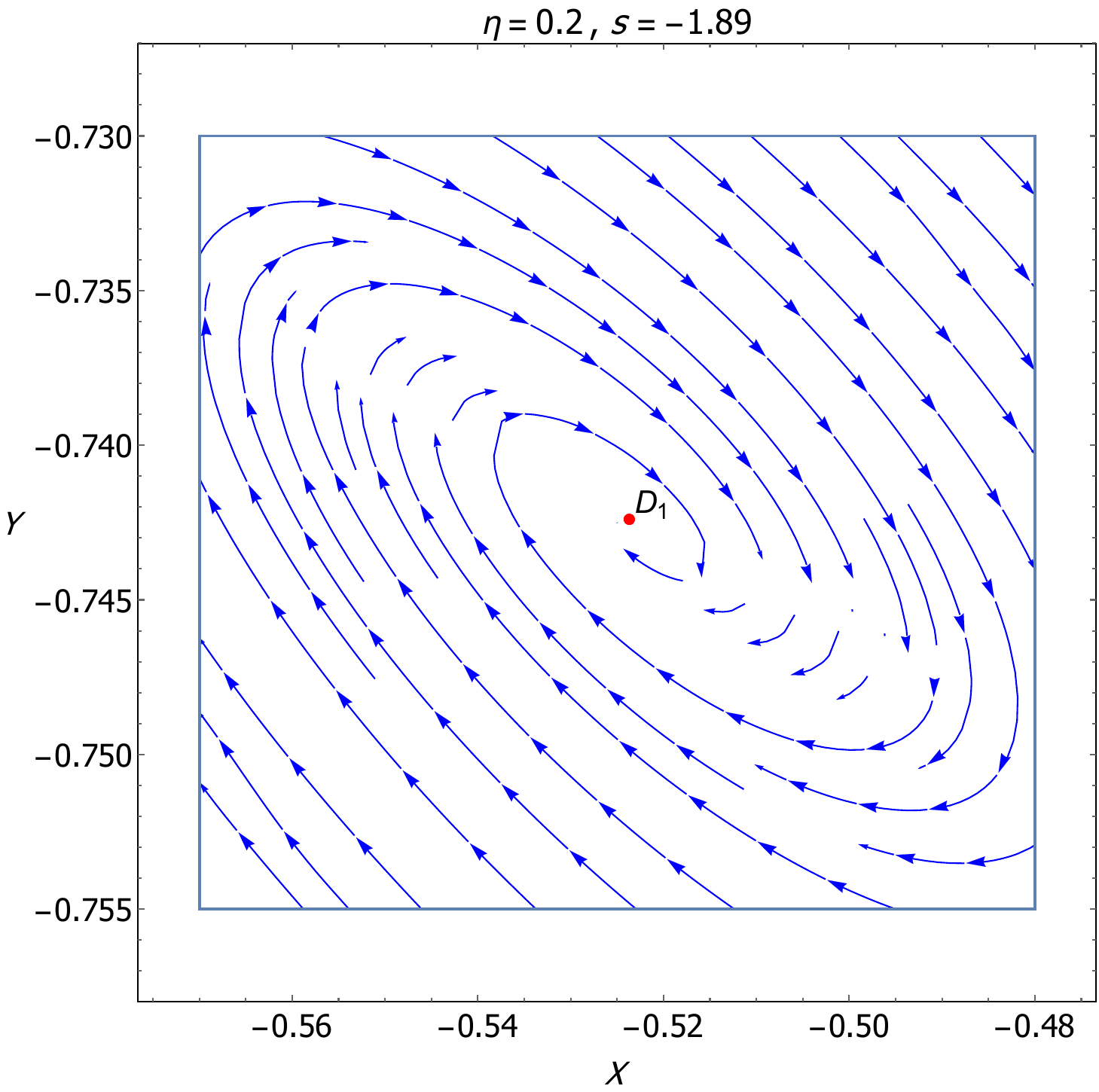}\label{D1stable}}
	\qquad
	\subfigure[]{%
		\includegraphics[width=7cm,height=6cm]{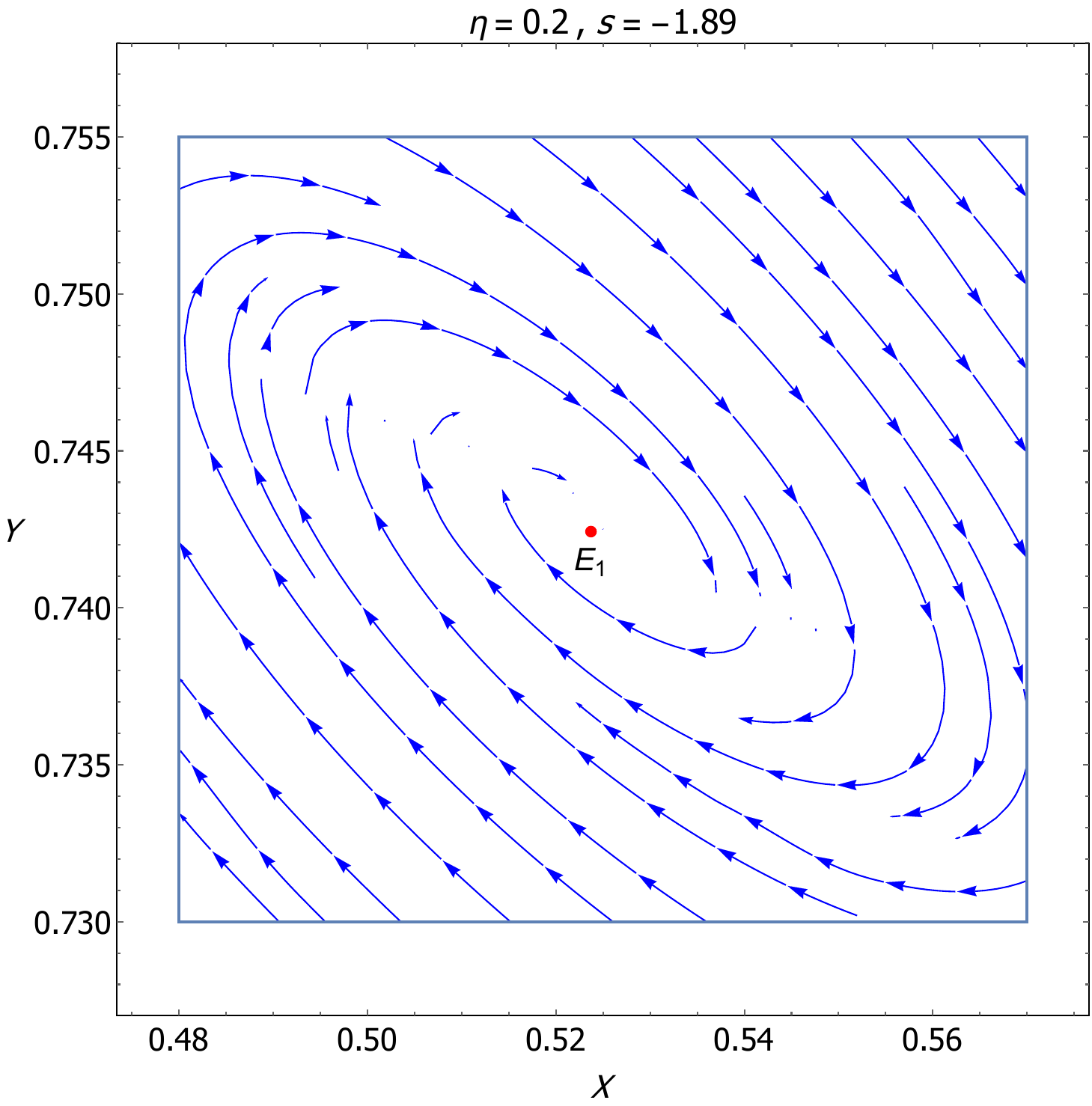}\label{E1stable}}
	\qquad
	\subfigure[]{%
		\includegraphics[width=10cm,height=6cm]{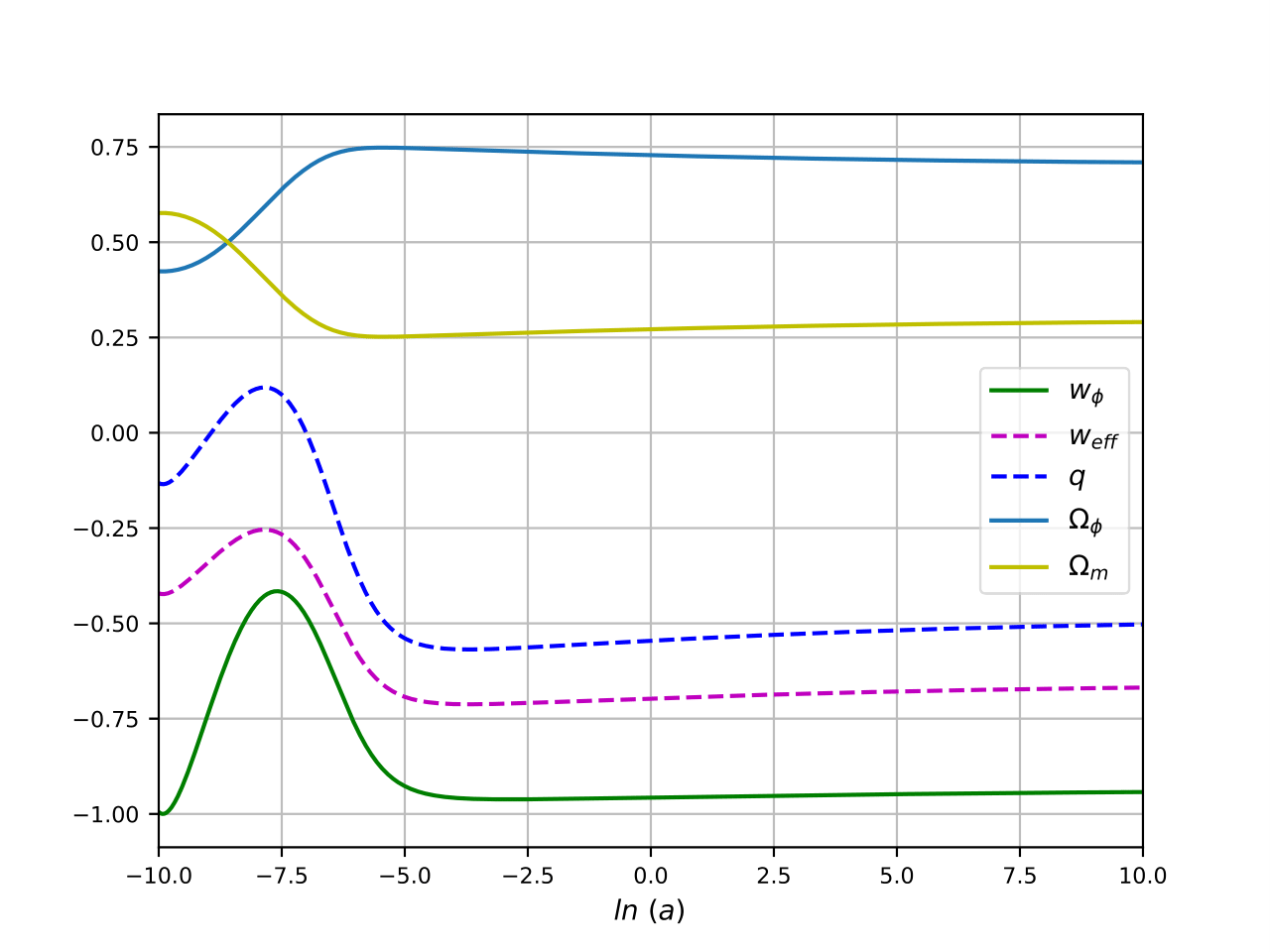}\label{Evolution_E}}
	\caption{In (a) and (b) trajectories show that late time scaling solutions $D_1$ and $E_1$ are attracted in quintessence era for $\eta=0.2, s=-1.89$. Evolution of cosmological parameters for the critical points are in panel (c) for $\eta=0.33, s=2.8$ showing that the late time accelerated solutions attracted in quintessence era.}
	\label{phasespace-figure-EF}
\end{figure}

\begin{figure}
	\centering
	\subfigure[]{%
		\includegraphics[width=7cm,height=6cm]{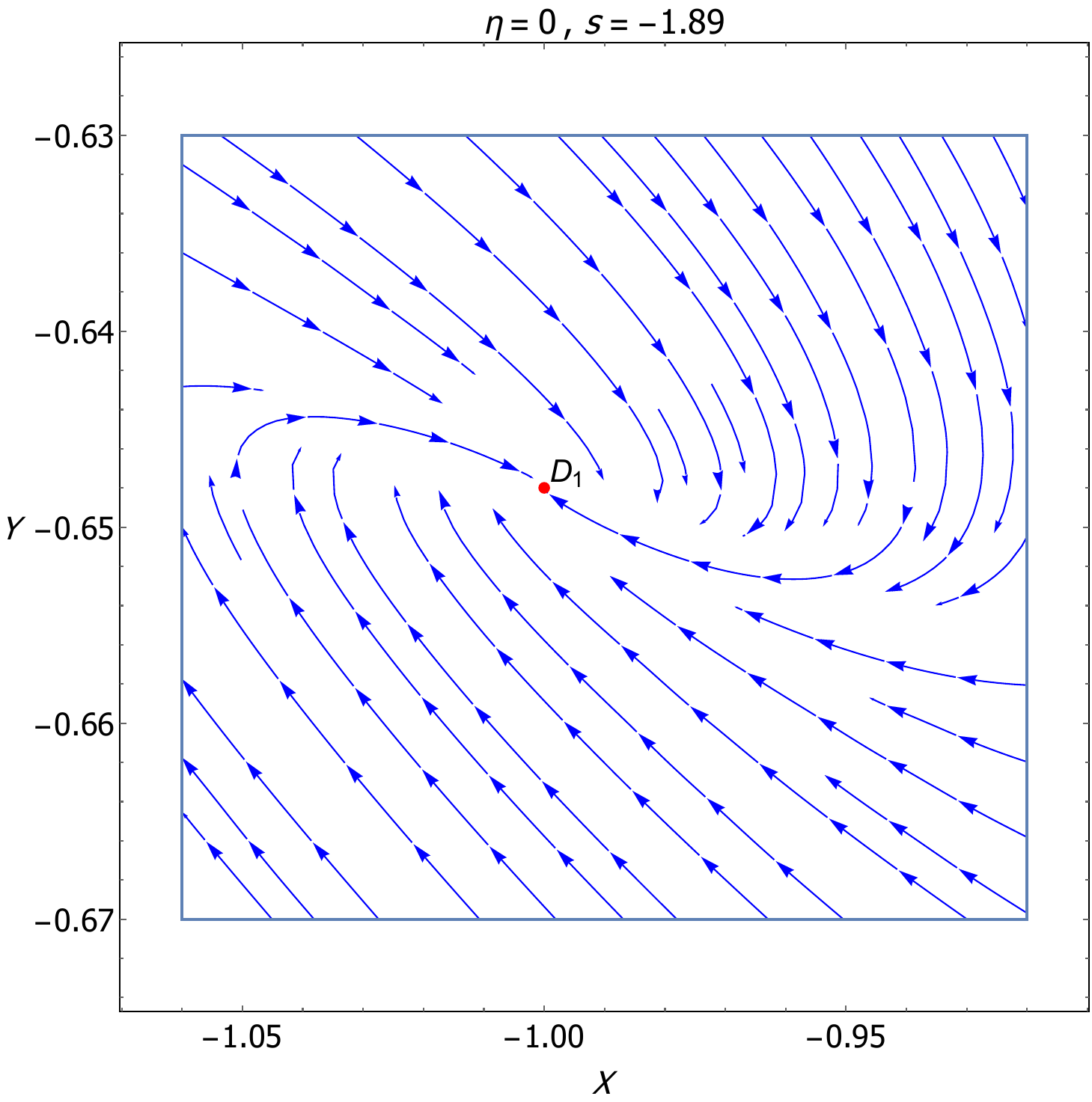}\label{Estable}}
	\qquad
	\subfigure[]{%
		\includegraphics[width=7cm,height=6cm]{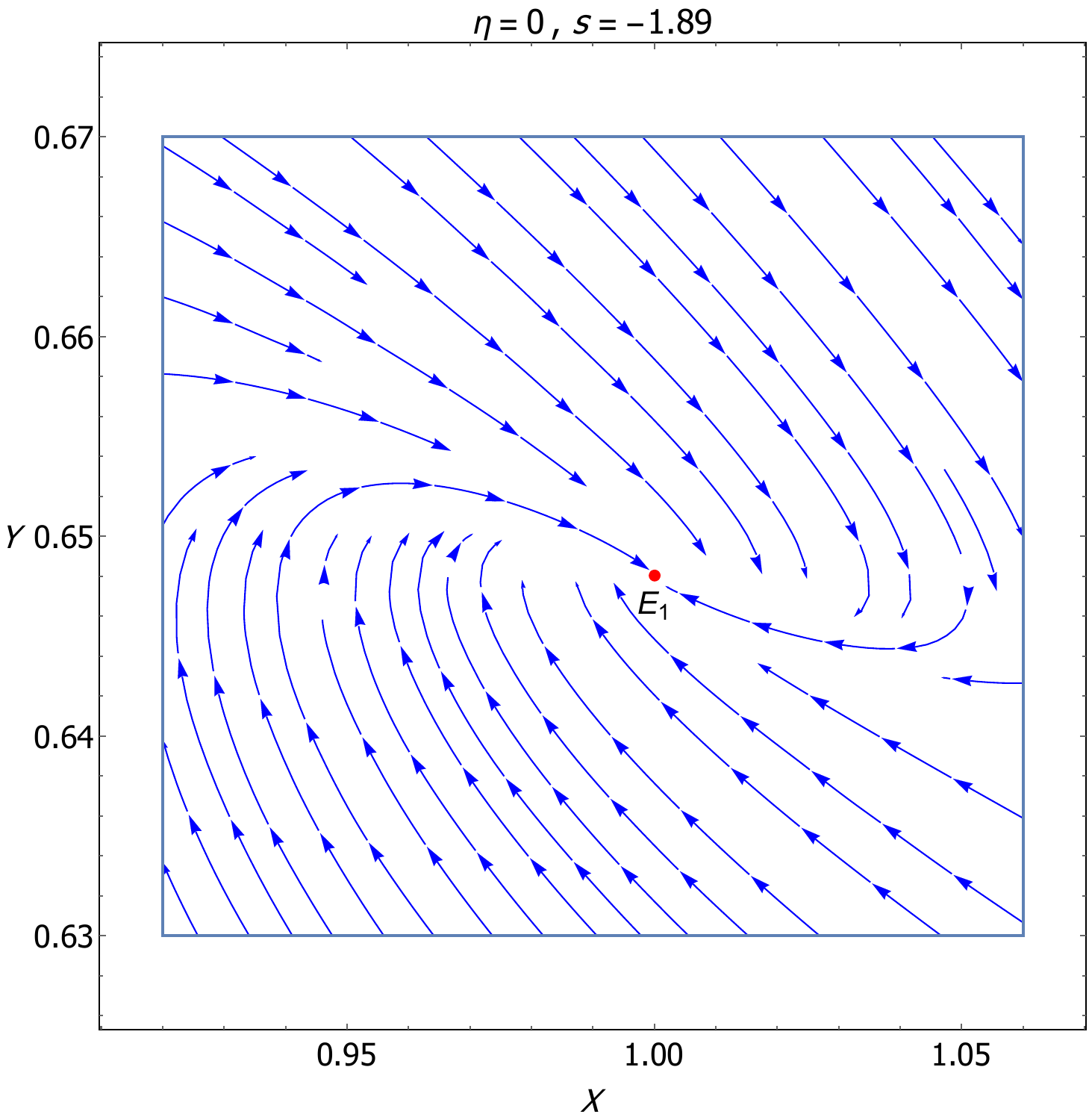}\label{Fstable}}
	\caption{Trajectories shows that late time decelerated scaling solutions $D_1$ ,$E_1$ are attracted in quintessence era for $\eta=0, s=-1.89$. }
	\label{FstableDeceleration}
\end{figure}


\begin{table}[ht] \centering \fontsize{10pt}{5pt}
	\caption{The critical points and their corresponding physical parameters (with their observational bounds) for the interaction model $ Q1=6\eta H\rho_{m}$ for power-law potential are presented.}
	\setlength{\tabcolsep}{0.05cm}
	\renewcommand{\arraystretch}{}
	\begin{tabular}
		[c]{|c|c|c|c|c|c|p{6cm}|}\hline
		\textbf{Critical} &$\mathbf{\Omega_{m}}$& $\mathbf{\Omega_{\phi}}$ &
		$\mathbf{\omega_{\phi}}$ & $\mathbf\omega_{eff}$ & $q$ & ~~~~~$-0.56<q<-0.49$  at late-time  \\
		\textbf{Points} &  &  &
		&  &  &  \\
		\hline\hline
		$A_{1}  $ & $1-y_{c}^2$ & $y_{c}^2$ &
		$-1$ & $-y_{c}^2$ & $\frac{1}{2}-\frac{3 y_{c}^2}{2}$ & ~~~~~~~~$ 0.812<y_c<0.841$  $\mbox{and}~ \eta >0$ \\
		\hline
		$B_{1}  $ & $0$ & $1$ &
		$\frac{s^2}{3}-1$ & $\frac{s^2}{3}-1$ & $\frac{s^2}{2}-1$ &    
		\multicolumn{1}{p{6.5cm}|}{\raggedright $ \bigg[\eta \leq 0.33 $ and $ (-1.01<s<-0.938 $ or $ 0.938<s<1.01)\bigg] $ or $ \bigg[0.33<\eta <0.353$ and $ (-\sqrt{3-6\eta}<s<-0.938$ or $0.938<s<\sqrt{3-6\eta })\bigg]  $}  \\
		\hline
		$C_{1} $ & $0$ & $1$ &
		$\frac{s^2}{3}-1$ & $\frac{s^2}{3}-1$ & $\frac{s^2}{2}-1$ & \multicolumn{1}{p{6.5cm}|}{\raggedright $\bigg[\eta \leq 0.33 $ and $ (-1.01<s<-0.938 $ or $ 0.938<s<1.01)\bigg] $ or $ \bigg[0.33<\eta <0.353$ and $ (-\sqrt{3-6\eta}<s<-0.938$ or $0.938<s<\sqrt{3-6\eta })\bigg] $} \\
		\hline
		$D_{1}  $ & $\frac{(1-2\eta)(s^2 +6\eta-3)}{s^2}$ & $\frac{2\eta s^2+3(2\eta-1)^2}{s^2}$ & $-\frac{2\eta s^2}{2\eta s^2 +3(2\eta-1)^2} $ & $-2\eta$ & $\frac{1}{2}-3\eta$ & \multicolumn{1}{p{6.5cm}|}{\raggedright $ \bigg[ 0.33<\eta <0.353$ and $(-\frac{\sqrt{3}}{2} \sqrt{\frac{4 \eta-12 \eta ^2+1}{\eta }}<s<-\sqrt{3-6\eta}$ or $\sqrt{3-6\eta}<s<\frac{\sqrt{3}}{2} \sqrt{\frac{4 \eta-12 \eta ^2+1}{\eta }})\bigg] $}  \\
		\hline
		$E_{1}  $ & $\frac{(1-2\eta)(s^2 +6\eta-3)}{s^2}$ & $\frac{2\eta s^2+3(2\eta-1)^2}{s^2}$ & $-\frac{2\eta s^2}{2\eta s^2 +3(2\eta-1)^2}$ & $-2\eta$ & $\frac{1}{2}-3\eta$ & \multicolumn{1}{p{6.5cm}|}{\raggedright $ \bigg[ 0.33<\eta <0.353$ and $(-\frac{\sqrt{3}}{2} \sqrt{\frac{4 \eta-12 \eta ^2+1}{\eta }}<s<-\sqrt{3-6\eta}$ or $\sqrt{3-6\eta}<s<\frac{\sqrt{3}}{2} \sqrt{\frac{4 \eta-12 \eta ^2+1}{\eta }})\bigg] $}  
		\\
		\hline\hline
	\end{tabular}
	\label{Physical parameters model1}
\end{table}
%


\subsection{Interaction model 2 : $ Q\propto\rho_{_{m}}$ }\label{}

Now, in this sub-section we shall discuss another interaction model in which the term $Q$ is chosen locally, i.e., not depending directly on Hubble parameter $H$ as \cite{Bohmer Interation2a,Bohmer Interaction2b} 
\begin{equation}\label{Interaction2}
	Q=\Gamma\rho_{m},
\end{equation}
where $\Gamma$ stands for a constant parameter which for $\Gamma>0$ indicating the flow of energy from DE to DM
and reverse for $\Gamma<0$.
Using the interaction term (\ref{Interaction2}) in the system (\ref{Gen_autonomous}) with exponential form of potential and coupling function of scalar field, i.e., 
$V(\phi)=V_{0} e^{(\beta H_{0}\phi)}$,  $\lambda(\phi)=\lambda_{0} e^{(\alpha H_{0}\phi)}$
(letting $\gamma=\frac{\Gamma}{H_{0}}$) and introducing a new dynamical variable
\begin{equation}\label{variables}
	r=\frac{H_{0}}{H+H_{0}},~~~~~~ \mbox{where}~H_{0}~\mbox{is}~\mbox{constant}
\end{equation}
we obtain a constraint equation as:
	\begin{equation}\label{constraint}
		y^{\alpha}=\sqrt{\frac{V_{0}^{\alpha}}{\lambda_{0}^{\beta}}}\left( \sqrt{3}H_{0} \left( \frac{1}{r}-1\right) \right) ^{\beta-\alpha} z^{\beta}
	\end{equation},
	
	and the following 4D autonomous system:  
\begin{eqnarray}
	\begin{split}
		\frac{dx}{dN}& = -3x-\frac{\alpha}{\sqrt{2}}\frac{x^2r}{(1-r)}-\frac{\beta}{\sqrt{2}}\frac{y^2r}{z^2(1-r)}-\frac{\gamma}{2}\frac{r}{(1-r)}\frac{(1-x^2z^2-y^2)}{xz^2},& \\
		\frac{dy}{dN}& = y\left[ \frac{\beta}{\sqrt{2}}\frac{r}{(1-r)} x+\frac{3}{2}\left( 1+x^2z^2-y^2 \right)\right],&\\
		\frac{dz}{dN}& = z\left[ \frac{\alpha}{\sqrt{2}}\frac{r}{(1-r)} x+\frac{3}{2}\left( 1+x^2z^2-y^2 \right) \right].&\\
		\frac{dr}{dN}& = \frac{3}{2}r(1-r)\left( 1+x^2z^2-y^2\right),~\label{auto-Exp-Int-20}
	\end{split}
\end{eqnarray} \\
where $\frac{\lambda^{'}}{\lambda} \frac{1}{H_{0}}=\alpha=$ constant and $\frac{V^{'}}{V} \frac{1}{H_{0}}=\beta=$ constant.
Now, the system (\ref{auto-Exp-Int-20}) has singularities at $x=0, z=0$ and $r=1$ respectively. So, we  multiply the right hand side(r.h.s.) of (\ref{auto-Exp-Int-20}) by $x^{2}z^2(1-r)$. Then the system (\ref{auto-Exp-Int-20}) reduces to the form :
\begin{eqnarray}
	\begin{split}
		\frac{dx}{dN}& = -3x^3z^2(1-r)-\frac{\alpha}{\sqrt{2}}x^4z^2r-\frac{\beta}{\sqrt{2}}x^2y^2r-\frac{\gamma}{2}xr(1-x^2z^2-y^2),& \\
		\frac{dy}{dN}& = x^2yz^2\left[ \frac{\beta}{\sqrt{2}}r x+\frac{3}{2}(1-r)\left( 1+x^2z^2-y^2 \right) \right],&\\
		\frac{dz}{dN}& = x^2z^3\left[ \frac{\alpha}{\sqrt{2}}r x+\frac{3}{2}(1-r)\left( 1+x^2z^2-y^2\right) \right].&\\
		\frac{dr}{dN}& = \frac{3}{2}x^2z^2r(1-r)^2\left( 1+x^2z^2-y^2 \right)~\label{auto-Exp-Int-21}
	\end{split}
\end{eqnarray}
We present the critical points and the corresponding physical parameters  in the table \ref{modelQ2EXP}.
The following are the critical points for the system (\ref{auto-Exp-Int-21}):
\begin{itemize}
	\item  I. Set of critical points : $ A_{2} = (0,y_{c},z_{c},r_{c}) $
	\item  II. Set of critical points : $ B_{2} = (0,1,z_{c},r_{c}) $
	\item  III. Set of critical points : $ C_{2} = (0,-1,z_c,r_c) $
	\item  IV. Set of critical points : $ D_{2} =  (x_c,y_{c},0,0) $
	\item  V. Set of critical points: $ E_{2} = \left(x_c,\frac{\sqrt\gamma}{\sqrt{\gamma-\sqrt{2}\beta x_c}},0,r_c\right) $
	\item  VI. Set of critical points : $ F_{2} = \left(x_c,-\frac{\sqrt\gamma}{\sqrt{\gamma-\sqrt{2}\beta x_c}},0,r_c \right) $
\end{itemize}


\begin{table}[ht] \centering \fontsize{10pt}{5pt}

		\caption{The critical points and the corresponding existence conditions are presented.}
		\setlength{\tabcolsep}{0.4cm}
		\renewcommand{\arraystretch}{}
		\begin{tabular}
			[c]{|c|c|p{6cm}|}\hline
			\textbf{Critical Points : }$(\mathbf{x,y,z,r})$ &
			$\textbf{Existence conditions} $ \\\hline\hline
			
			$A_{2} : (0,y_{c},z_{c},r_{c}) $ & \multicolumn{1}{p{10cm}|}{\raggedright 
				(i)~$\alpha=\beta=0$ and $r_{c}\neq 0$ and $0\leq y_{c}^{2}\leq1$~,or \\
				(ii)~$\alpha=\beta$ and $r_{c}\neq 0$ and $y_{c}=z_{c}$ and $V_{0}=\lambda_{0}$ and $0\leq y_{c}^{2}\leq 1$.}  \\
			\hline	
			$B_{2} : (0,1,z_{c},r_{c})$ & \multicolumn{1}{p{10cm}|}{\raggedright (i)~$\alpha=\beta=0$ and $r_{c}\neq 0$ ~,or \\
				(ii)~$\alpha=\beta$ and $r_{c}\neq 0$ and  $V_{0}=\lambda_{0}$ and $z_{c}= 1.$} \\
			\hline
			$C_{2} : (0,-1,z_c,r_c)$ &
			\multicolumn{1}{p{10cm}|}{\raggedright (i)~$\alpha=\beta=0$ and $r_{c}\neq 0$ ~,or \\
				(ii)~$\alpha=\beta$ and $r_{c}\neq 0$ and  $V_{0}=\lambda_{0}$ and $z_{c}= 1$ and $\alpha$ be even.} \\
			\hline
			$D_{2} :  (x_c,y_{c},0,0) $ & \multicolumn{1}{p{10cm}|}{\raggedright does not exist as $y_{c}$ be any in Eq.(\ref{constraint})}  \\
			\hline
			
			$E_{2} : \left(x_c,\frac{\sqrt\gamma}{\sqrt{\gamma-\sqrt{2}\beta x_c}},0,r_c\right)$ &
			\multicolumn{1}{p{10cm}|}{\raggedright 
				(i)~$\beta>0$ and $\beta>\alpha$ and $r_{c}\neq0$ and $x_{c}\neq0$ and $\gamma=0$~,or \\
				(ii)~$\beta>0$ and $\beta<\alpha$ and $r_{c}\neq1$ and $x_{c}\neq0$ and $\gamma=0$}  \\
			\hline
			$ F_{2}:\left(x_c,-\frac{\sqrt\gamma}{\sqrt{\gamma-\sqrt{2}\beta x_c}},0,r_c \right)$ &
			\multicolumn{1}{p{10cm}|}{\raggedright (i)~$\beta>0$ and $\beta>\alpha$ and $r_{c}\neq0$ and $x_{c}\neq0$ and $\gamma=0$~,or \\
				(ii)~$\beta>0$ and $\beta<\alpha$ and $r_{c}\neq1$ and $x_{c}\neq0$ and $\gamma=0$}   \\
			
			\hline\hline
			
		\end{tabular}
		\label{existence condition}
	
\end{table}


\begin{table}[ht] \centering \fontsize{10pt}{5pt}
	\caption{The critical points and their corresponding physical parameters for the interaction model $ Q2=\Gamma\rho_{m} $ for exponential potential are presented.}
	
	\begin{tabular}	[c]{|c|c|c|c|c|c|c|}\hline
		\textbf{Critical Points} & $(\mathbf{x,y,z,r})$ &$\mathbf{\Omega_{m}}$ &
		$\mathbf{\Omega_{\phi}}$ &~~~~ $\mathbf{\omega_{\phi}}$ & $ \mathbf{\omega_{eff}} $ &~~~~~ $q$ 
		\\\hline\hline
		$ A_{2} $ & $ (0,y_{c},z_{c},r_{c}) $ & $1-y^2_c$ &
		$y_{c}^2$ & $-1$ & $-y_{c}^2$ &~~ $\frac{1}{2}-\frac{3y^2_c}{2}$\\\hline
		$ B_{2} $ & $ (0,1,z_{c},r_{c}) $ & $ 0 $ &
		$ 1 $ & $ -1 $ & $ -1 $ &~~ $-1$ \\\hline
		$ C_{2} $ & $ (0,-1,z_c,r_c) $ & $ 0 $ &
		$ 1 $ & $ -1 $ & $ -1 $ &~~ $-1$\\\hline
		$ D_{2} $ & $ (x_c,y_{c},0,0) $   & $ 1-y^2_c $ & $ y_{c}^2 $ & $ -1$ & $ -y_{c}^2 $ &~~ $\frac{1}{2}-\frac{3y^2_c}{2}$ \\\hline
		$ E_{2} $ & $ (x_c,\frac{\sqrt\gamma}{\sqrt{\gamma-\sqrt{2}\beta x_c}},0,r_c) $ & $ \frac{\sqrt{2}\beta x_c}{\sqrt{2}\beta x_c -\gamma} $ & $ \frac{\gamma}{-\sqrt{2}\beta x_c +\gamma} $ & $-1 $ & $ \frac{\gamma}{\sqrt{2}\beta x_c -\gamma} $ &~~$\frac{\sqrt{2}\beta x_c+2\gamma}{2\sqrt{2}\beta x_c -2\gamma}$ \\\hline
		$ F_{2} $ & $ (x_c,-\frac{\sqrt\gamma}{\sqrt{\gamma-\sqrt{2}\beta x_c}},0,r_c) $ & $   \frac{\sqrt{2}\beta x_c}{\sqrt{2}\beta x_c -\gamma} $ & $ \frac{\gamma}{-\sqrt{2}\beta x_c +\gamma} $ & $ -1 $ & $ \frac{\gamma}{\sqrt{2}\beta x_c -\gamma} $  &~~$\frac{\sqrt{2}\beta x_c+2\gamma}{2\sqrt{2}\beta x_c -2\gamma}$
		\\\hline\hline
	\end{tabular}
	\label{modelQ2EXP}
\end{table}%
%

\begin{figure}
	\centering
	\includegraphics[width=8cm,height=6cm]{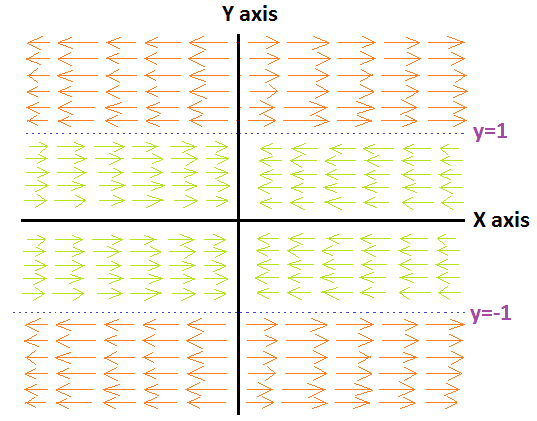}
	\caption{ The vector field (projected on $x-y$ plane) near $A_2$ is topologically equivalent to this figure in the cosmologically viable region for parameter value $r_c\gamma>0$. The red arrow region represents unstable vector field and the green arrow region represents the stable vector field  }
	\label{vector field A_2}
\end{figure}
\begin{figure}
	\centering
	\includegraphics[width=8cm,height=6cm]{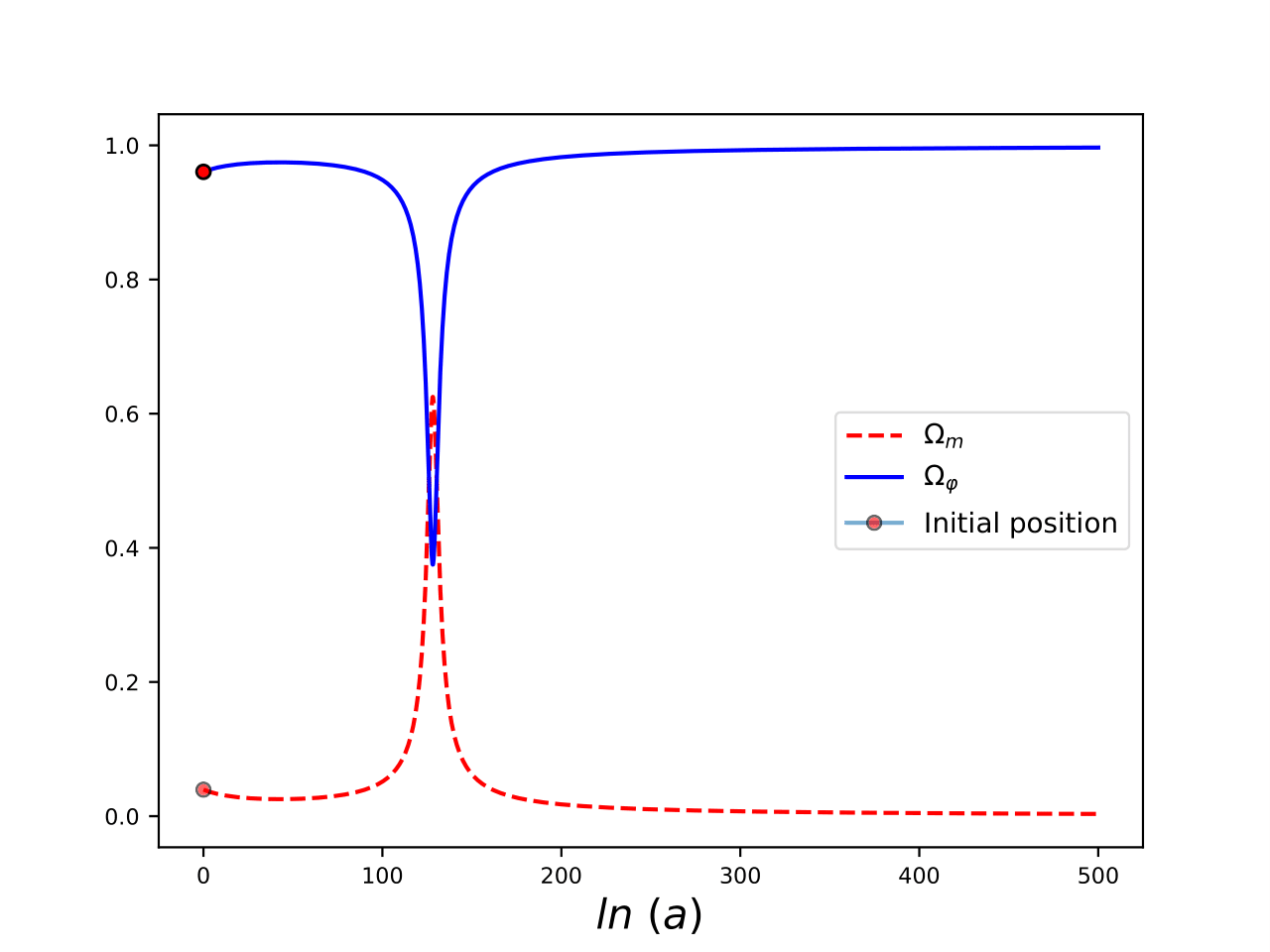}
	\caption{Evolution of energy density parameters for DE and DM, when initial conditions are chosen near the set of points $A_2$ for the parameter values $\alpha = \beta = \gamma =1$.  }
	\label{A_2 evolution}
\end{figure}
\subsubsection{Phase space analysis for interaction 2}
We shall discuss the phase space analysis of sets of critical points arise from the 4D autonomous system. 
\begin{itemize}
	
	\item
	Set of critical points $A_2$ exists for all parameters $\alpha, \beta, \gamma$. The set corresponds to a scaling solution in phase space. Dark energy associated with the set behaves as cosmological constant ($\omega_{\phi}=-1$). Accelerated evolution of the universe ($\omega_{eff}\longrightarrow -1$) is achieved by the set of points when it becomes DE dominated for $y_c\longrightarrow 1$. On the other hand, for $y_c\longrightarrow 0$ the set describe DM dominated decelerated universe. Evolution of the cosmological parameters are shown in figure \ref{A_2 evolution} for the parameter values $\alpha = \beta = \gamma =1$.  Eigenvalues of perturbation matrix are : $\left\{\mu_1=-\frac{r_{c}\gamma}{2}\left( 1-y_{c}^{2}\right) ,~\mu_2=0,~\mu_3=0,~\mu_4=0   \right\}$. 
	
	\item
	Set of critical points $B_2$ and $C_2$ are same in all respects in the cosmological point of view. These sets represent completely potential energy of scalar field dominated solutions where DE behaves as cosmological constant ($\omega_{\phi}=-1$). There exists accelerating universe always near the set of critical points ($\omega_{eff}=-1, q=-1$). Eigenvalues for the sets $B_2$ and $C_2$ are: $\left\{\mu_1=0,~\mu_2=0,~\mu_3=0,~\mu_4=0   \right\}$.
	These sets are non-hyperbolic in nature but not normally hyperbolic set.
	
	It is to be noted that $B_2$ and $C_2$ are special case of $A_2$ when $y_c=\pm 1$.  So the stability near $B_2$ and $C_2$ are regulated by the eigenvalue $(\mu_1)$ corresponding to $X-$ variable.\\
	When $y_c\neq \pm 1$, $\mu_1=-\frac{r_c\gamma}{2}(1-y_c^2)\neq 0$ and $r_c\gamma>0$ in the cosmological viable region.  In this context one may note that for $|y_c|>1$, $\mu_1$ is strictly positive ($\mu_1>0$) and for $|y_c|<1$, $\mu_1$ is strictly negative ($\mu_1<0$).  So the vector field bifurcates from unstable to stable in nature when $y_c$ crosses $1$ from larger to smaller value through $1$ and smaller to larger value through $-1$(see figure \ref{vector field A_2}).
	
	
	\item
	Set of critical points $D_2$ corresponds to a scaling solution behaves similarly to the set of points $A_2$ in the phase space. It is also a non-hyperbolic type set since the eigenvalues at the set of points are: $\left\{\mu_1=0,~\mu_2=0,~\mu_3=0,~\mu_4=0   \right\}$. As it is not appropriate to apply Hartman-Grobman theorem or center manifold theory for the case of no non-zero eigenvalue, so stability analysis of the set of points $D_2$ is found to be complicated.
	
	\item
	Sets of critical points $E_2$ and $F_2$ are same in all respects. They correspond to DE-DM scaling solutions in the phase space. Depending on some parameter restrictions, the sets show the accelerating universe. Eigenvalues of perturbation matrix at the set of points $E_2$ and $F_2$ are: $\left\{\mu_1=\frac{\sqrt{2} \beta\gamma x_{c} r_c}{2(\sqrt{2}\beta x_c -\gamma)},~\mu_2=0,~\mu_3=0,~\mu_4=0   \right\}$ which indicates the sets are non-hyperbolic in nature and to characterize these points linear stability theory or Hartman-Grobman theorem is not appropriate to apply.  So we apply center manifold theory.
\end{itemize}
\subsection{The speed of sound and Classical stability analysis}

 The authors in Refs. \cite{Dutta2018,Basilakos2019,Khyllep2022} studied the dynamical analysis in order to investigate the scalar field dark energy model at the perturbation level where
 adiabatic sound speed $(C_{s})$ has an important role to investigate the classical stability as well as causality of the model. The square of speed of sound is defined as $C_{s}^{2}=\frac{\partial p_\phi}{\partial \rho_{\phi}}$ which appears as a coefficient of the term $\frac{k^{2}}{a^{2}}$ ( where $k$ is the co-moving momentum and $a$ is the usual scale factor). Note that the classical fluctuations are considered to be stable when $C_{s}^{2}$ takes positive ($C_{s}^{2}\geq 0$) values \cite{Biswas:2015cva,Mahata:2013oza,S.Kr.Biswas2015a,Piazza}. On the other hand, it violates the causality when speed of sound diverges, i.e., for $C_{s}^{2}>1$ and ghost instabilities may be occurred for $C_{s}^{2}<0$. We shall now investigate the classical stability for the critical points of the interaction model 1 as well as Interaction model 2.

\subsubsection{For interaction 1}

To find the squared sound speed, we first eliminate the term $\frac{1}{2}\lambda \dot{\phi}^2$ from (\ref{energy density scalar}) and (\ref{pressure scalar}) to obtain
$$p_{\phi}=\rho_{\phi}-2V.$$
From the definition, we have

$$C_{s}^{2}=\frac{\partial p_\phi}{\partial \rho_{\phi}}=1-2\frac{d V}{d \phi} \frac{\dot{\phi}}{\dot{\rho_{\phi}}}$$
Putting the value of $\dot{\rho_{\phi}}$ from Eqn. (\ref{continuity-scalar}) and dividing both the numerator and denominator of the last term of the above by $\dot{\phi}$, one gets
$$C_{s}^{2}=1+\frac{2V'}{\frac{Q}{\dot{\phi}}+3H\lambda \dot{\phi}}$$
By using the expression of $Q$ in (\ref{Interaction1}), the above equation (after dividing both numerator and denominator of last term of the above by $3\sqrt{3}H^3$) takes the form in terms of dynamical variables and free parameters:
$$C_{s}^{2}=1+\frac{2sy^3}{\frac{\sqrt{6}\eta}{x}(1-x^2 z^2 -y^2)+\sqrt{6}z^2 x}$$
Finally, for model 1 (where we have considered a 2D system after replacing $z$ by $y$) the explicit form of sound speed  in terms of dynamical variables and model parameters is as follows:

\begin{equation}
	C_{s}^{2}=1+\sqrt{\frac{2}{3}}\frac{ s x y^3}{\eta  \left(1-x^2 y^2-y^2\right)+x^2 y^2}.
\end{equation}

The critical points and the conditions for their classical stability are shown in the table \ref{classical stability modelQ1}. 

\begin{table}[ht] \centering \fontsize{10pt}{5pt}
	\caption{The critical points, square of sound speed square and the classical stability conditions are presented.}
	\setlength{\tabcolsep}{0.25cm}
	\renewcommand{\arraystretch}{}
	\begin{tabular}
		[c]{|c|c|c|p{6cm}|}\hline
		\textbf{Critical Points : }$(\mathbf{x,y})$& $\mathbf{C_{s}^{2}}$ &
		$\textbf{Classical stability} (C_{s}^{2}\geq0)$ \\\hline\hline
		$A_{1}:(0,y_{c})$ &  $ 1
		$ &
		for any $\eta,~s$ \\ \hline
		
		$B_{1}:\left(-\frac{s}{\sqrt{6-s^2}},~~ \sqrt{1-\frac{s^2}{6}}\right)$ & $ \frac{1}{3} \left(s^2-3\right) $ & $-\sqrt{6}<s\leq -\sqrt{3}~\mbox{or}~ \sqrt{3}\leq s<\sqrt{6}$ \\
		\hline
		$C_{1}:\left(\frac{s}{\sqrt{6-s^2}},~~ -\sqrt{1-\frac{s^2}{6}}\right)$ & $\frac{1}{3} \left(s^2-3\right)  $ &
		$ -\sqrt{6}<s\leq -\sqrt{3}~\mbox{or}~ \sqrt{3}\leq s<\sqrt{6} $  \\
		\hline
		$D_{1}:\left(\frac{\sqrt3 (2\eta-1)}{\sqrt{4\eta s^2 +3(2\eta-1)^2}},~~\frac{\sqrt{4\eta s^2 +3(2\eta-1)^2}}{\sqrt{2} s}\right)$ & $ -\frac{2 \eta  s^2}{2 \eta  \left(6 \eta +s^2-6\right)+3}  $ & \multicolumn{1}{p{7cm}|}{\raggedright $ \bigg[  -\frac{1}{2}<\eta <0 $ and 
			$(-\frac{\sqrt{3}}{2}  \sqrt{4-4 \eta -\frac{1}{\eta }}<s\leq -\sqrt{3-6 \eta }$ or $\sqrt{3-6 \eta }\leq s<\frac{\sqrt{3}}{2}  \sqrt{4-4 \eta -\frac{1}{\eta }}) \bigg]  $ or $\bigg[ \eta =0$ and $ (s+\sqrt{3}\leq 0$ or $s\geq \sqrt{3}) \bigg] $}  \\
		\hline
		
		$E_{1}:\left(-\frac{\sqrt3 (2\eta-1)}{\sqrt{4\eta s^2 +3(2\eta-1)^2}},~~-\frac{\sqrt{4\eta s^2 +3(2\eta-1)^2}}{\sqrt{2} s} \right)$& $ -\frac{2 \eta  s^2}{2 \eta  \left(6 \eta +s^2-6\right)+3} $ &
		\multicolumn{1}{p{7cm}|}{\raggedright $\bigg[-\frac{1}{2}<\eta <0 $ and 
			$( -\frac{\sqrt{3}}{2}  \sqrt{4-4 \eta -\frac{1}{\eta }}<s\leq -\sqrt{3-6 \eta }$ or $\sqrt{3-6 \eta }\leq s<\frac{\sqrt{3}}{2}  \sqrt{4-4 \eta -\frac{1}{\eta }}  )   \bigg]$ or $\bigg[ \eta =0$ and $ ( s+\sqrt{3}\leq 0$ or $s\geq \sqrt{3} )  \bigg] $}  \\
		
		\hline\hline
	\end{tabular}
	\label{classical stability modelQ1}
\end{table}


\subsubsection{For interaction 2}	
For the  Interaction 2, the expression of the sound speed will take the form
\begin{equation}
	C_{s}^{2}=1+\frac{2 \sqrt{2} \beta  r x y^2}{\gamma  r \left(1-x^2 z^2-y^2\right)+6 x^2 z^2}.
\end{equation}

and the conditions for classical stability are presented in the following table \ref{classical stability modelQ2}.

\begin{table}[ht] \centering \fontsize{10pt}{5pt}
	\caption{The critical points, sound speed square and the classical stability conditions are presented.}
	\setlength{\tabcolsep}{0.3cm}
	\renewcommand{\arraystretch}{}
	\begin{tabular}
		[c]{|c|c|c|p{6cm}|}\hline
		\textbf{Critical Points}&$(\mathbf{x,y,z,r})$& $\mathbf{C_{s}^{2}}$ &
		$\textbf{Classical stability} (C_{s}^{2}\geq0)$ \\\hline\hline
		$A_{2}$ & $(0,y_{c},z_{c},r_{c})$ &  $ 1
		$ &
		for any $\alpha,\beta,\gamma$ \\ \hline
		
		$B_{2}$ & $(0,1,z_{c},r_{c})$ & $ Indeterminate $ & $Undefined$ \\
		\hline
		$C_{2}$ & $(0,-1,z_c,r_c)$ & $Indeterminate $ &
		$ Undefined $  \\
		\hline
		$D_{2}$ & $(x_c,y_{c},0,0) $ & $ Indeterminate $ &  $Undefined$  \\
		\hline
		
		$E_{2}$ & $ \left(x_c,\frac{\sqrt\gamma}{\sqrt{\gamma-\sqrt{2}\beta x_c}},0,r_c\right)$& $ -1 $ &
		$ Not ~stable $  \\
		\hline
		$F_{2}$ & $\left(x_c,-\frac{\sqrt\gamma}{\sqrt{\gamma-\sqrt{2}\beta x_c}},0,r_c \right)$& $ -1 $ &
		$ Not ~stable  $  \\
		
		\hline\hline
	\end{tabular}
	\label{classical stability modelQ2}
\end{table}



\newpage

\section{Center manifold theory}\label{CMT}

\begin{center}
	\textbf{Mathematical background}
\end{center}
We follow Perko \cite{perko} and Boehmer \cite{Boehmer}, for discussing the mathematical background of the center manifold theory. When Jacobian matrix corresponding to the given autonomous system at the critical point has zero eigenvalue(s), linear theory fails to
provide information on the stability of that critical point. In this case, use of centre manifold is interested because it reduce the dimension
of the system near that critical point so that stability of the reduced system can be investigated. There always exists an
invariant local center manifold $W^c$ passing through the fixed point to which the system could be restricted to study its
behaviour in the neighbourhood of the fixed point. The stability of the reduced system determines the stability
of the system at that point.\bigbreak
Let $u\in\mathbb{R}^c$ and $v\in\mathbb{R}^s$.  An arbitrary dynamical system with zero eigenvalues in the Jacobian matrix can always be written in the following form
\begin{eqnarray}
	\begin{split}
		\dot{u}&=Au+f(u,v),\\\dot{v}&=Bu+g(u,v),\label{cm1}
	\end{split}
\end{eqnarray}
where
\begin{eqnarray}
	\begin{split}
		&f(0,0)=0,~~Df(0,0)=0,\\
		&g(0,0)=0,~~Dg(0,0)=0.\label{cm2}
	\end{split}
\end{eqnarray}
Here we assume that the critical point is located at the origin and $Df$ denotes the matrix of first derivatives of the
vector valued function $f$. $A$ is a $c\times c$ matrix having eigenvalues with zero real parts and $B$ is an $s\times s$ matrix having
eigenvalues with non-zero real parts.\\
It is noted that a dynamical system with zero eigenvalues can always be rewritten into the above form by linear shifting and matrix coordinate transformations. We will show this construction explicitly when we discuss the scalar field model below.\bigbreak
\textbf{ Definition.}	We call the space
\begin{eqnarray}
	W^c(0)=\left\{(u,v)\in\mathbb{R}^c\times \mathbb{R}^s|v=h(u), |u|<\delta,h(0) = 0,Dh(0) = 0 \right\}
\end{eqnarray}
for $\delta$ sufficiently small, the center manifold for the system (\ref{cm1}).  \bigbreak
Next, we need to construct this center manifold explicitly. By differentiating the defining equation $v = h(u)$ with
respect to the independent variable, we get $\dot{v} = Dh(u)\dot{u}$ where we used the chain rule. Eliminating $\dot{u}$ and $\dot{v}$ via (\ref{cm1}),
one arrives and the following quasilinear partial differential equation which $h$ has to satisfy
\begin{align}
	\mathcal{N}(h(u)) = Dh(u) [Au + f (u, h (u))] - Bh(u) - g(u, h(u)) = 0,
\end{align}
and the flow on the center manifold $W^c(0)$ is defined by the
system of differential equations
\begin{align}
	\dot{u}=Cu+f(u,h(u))
\end{align}
for all $u\in \mathbb{R}^c$ with $|u|<\delta$.

\subsection{Stability analysis for normally hyperbolic critical points of interaction model 1}
\subsubsection{$~Critical~point~A_1$}

The Jacobian matrix at the critical point $A_1$ corresponding to the autonomous system (\ref{autonomous2D}) can be put as
\begin{equation}\renewcommand{\arraystretch}{1.5}	
	J(A_1)=\begin{bmatrix}
		3\eta (y_c^2-1) & 0 \\
		0  &  0 
	\end{bmatrix}.\label{eq6}	
\end{equation}	
The eigenvalues of $J(A_1)$ are $3\eta (y_c^2-1) $ and $0$ and the corresponding eigenvectors are $[1,0]^T$ and  $\left[0,1\right]^T$ respectively.  Since the critical point $A_1$ is non-hyperbolic in nature, so we use center manifold theory for analysing the stability of this critical point.  First we take the shifting transformation $x=X, y=Y+y_c$ for a fixed $y_c$ so that the critical point $A_1$ moves to the origin.
By center manifold theory there exists a continuously differentiable function 	$h:\mathbb{R}$$\rightarrow$$\mathbb{R}$ such that
\begin{align}
	X=h(Y)=a_1Y^2+a_2Y^3+higher~order~terms\label{eq9}
\end{align}
Differentiating both side of (\ref{eq9}) with respect to $N$, then we get
\begin{eqnarray}
	\frac{dX}{dN}=(2a_1Y+3a_2Y^2)\frac{dY}{dN}\label{eq10}
\end{eqnarray}	
We only concern about the non-zero coefficients of the lowest power terms in center manifold theory as we analyse arbitrary small neighbourhood of the origin, so it is not needed to determine the another coefficients.  By comparing the coefficients both sides of (\ref{eq10}) we get $a_1=0,~a_2=0$.  In fact $a_i=0$ for all $i\in \mathbb{N}$. Hence the equation of center manifold can be expressed as
\begin{align}
	X=0
\end{align}
and the flow on the center manifold near the origin is determined by	
\begin{align}
	\frac{dY}{dN}=0
	\label{flow A1}
\end{align}
It follows that the center manifold is completely lying on $Y$ axis but the flow on the center manifold can not be determined from flow equation.

\subsection{Stability analysis for non-hyperbolic sets of interaction model 2}	

\subsubsection{$~Set~of~critical~points~A_2$}	

The Jacobian matrix at the critical point $A_2$ corresponding to the autonomous system (\ref{auto-Exp-Int-21}) can be put as
\begin{equation}\renewcommand{\arraystretch}{1.5}
	J(A_2)=\begin{bmatrix}
		\frac{1}{2}\gamma r_c(y_c^2-1) & 0 & 0&0\\	
		0  &  0 & 0&0\\
		0 & 0 & 0&0\\
		0&0&0&0 
	\end{bmatrix}.\label{equ54}	
\end{equation}	
The eigenvalues of $J(A_2)$ are $\frac{1}{2}\gamma r_c(y_c^2-1)$, $0$, $0$ and $0$.  $[1,0,0,0]^T$ be the eigenvector corresponding to the eigenvalue $\frac{1}{2}\gamma r_c(y_c^2-1)$ and $\left[0,1,0,0\right]^T$, $[0,0,1,0]^T$ and $[0,0,0,1]^T$ are the eigenvectors corresponding to the eigenvalue $0$.  For a fixed $y_c$, $z_c$ and $r_c$, first we take the shifting transformation $x=X, y=Y+y_c, z=Z+z_c, r=R+r_c$ so that the critical point $A_2$ moves to the origin.  By center manifold theory there exists a continuously differentiable function 	$h$:$\mathbb{R}^3$$\rightarrow$$\mathbb{R}$ such that
\begin{align}
	X=h(Y,Z,R)=a_1Y^2+a_2Z^2+a_3R^2+a_4 YZ+a_5YR+a_6ZR+higher~order~terms\label{equ57}
\end{align}
Differentiating both side of (\ref{equ57}) with respect to $N$, then we get
\begin{eqnarray}\renewcommand{\arraystretch}{1.5}	
	\frac{dX}{dN}=\begin{bmatrix}
		2a_1Y+a_4Z+a_5R&~~ 2a_2Z+a_4Y+a_6R &~~ 2a_3R+a_5Y+a_6Z
	\end{bmatrix}\begin{bmatrix}
		\frac{dY}{dN}\\
		\frac{dZ}{dN}\\
		\frac{dR}{dN}
	\end{bmatrix}\label{equ58}
\end{eqnarray}	
By comparing the coefficient of lowest order terms, we can get $a_i=0$ for all $i\in \mathbb{N}$.  Hence, the center manifold can be defined as
\begin{align}
	X=0
\end{align}
and the flow on the center manifold is obtained by
\begin{align}
	Y'&=0\\
	Z'&=0\\
	R'&=0
\end{align}
It follows that the center manifold contains in the three dimensional space $YZR$ but the dynamics on the vector field can not be determined from the flow equations.

\subsubsection{$~Set~of~critical~points~E_2$}

\begin{figure}[tbp]
	\centering
	\includegraphics[width=0.4\textwidth]{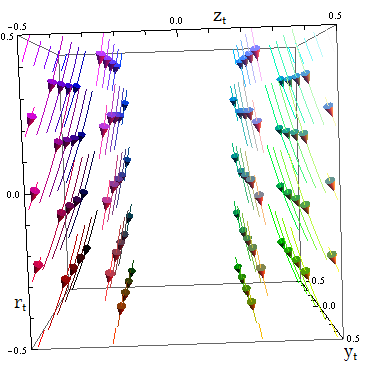}
	\caption{Dynamics on the center manifold is given locally (up to second order) by (\ref{eqn63}), (\ref{eqn64}) and (\ref{eqn65}).  For existence the domain of critical point and to satisfy the condition (\ref{eqn18}) we have drawn this $3$-dimensional phase plot for $x_c=-1,\beta=1,\gamma=2,r_c=2,\alpha=1$, which conclude that the origin is unstable in nature.}
	\label{space_of_critical_point}
\end{figure}

The Jacobian matrix at the critical point $E_2$ corresponding to the autonomous system (\ref{auto-Exp-Int-21}) can be put as
\begin{equation}\renewcommand{\arraystretch}{1.5}	
	J(E_2)=\begin{bmatrix}
		\frac{\sqrt{2}\beta\gamma r_c}{2(\sqrt{2}\beta x_c-\gamma)} & r_c\sqrt{\gamma}\sqrt{\gamma-\sqrt{2}\beta x_c} & 0&0\\	
		0  &  0 & 0&0\\
		0 & 0 & 0&0\\
		0&0&0&0 
	\end{bmatrix}.\label{eqn57}	
\end{equation}	
The eigenvalues of $J(E_2)$ are $\frac{\sqrt{2}\beta\gamma r_c}{2(\sqrt{2}\beta x_c-\gamma)}$, $0$, $0$ and $0$.  $[1,0,0,0]^T$ be the eigenvector corresponding to the eigenvalue $\frac{\sqrt{2}\beta\gamma r_c}{2(\sqrt{2}\beta x_c-\gamma)}$ and $\left[\frac{\sqrt{2}(\gamma-\sqrt{2}\beta x_c)^{\frac{3}{2}}}{\beta \sqrt{\gamma}},1,0,0\right]^T$, $[0,0,1,0]^T$ and $[0,0,0,1]^T$ are the eigenvectors corresponding to the eigenvalue $0$.  For a fixed $x_c$ and $r_c$, first we take the shifting transformation $x=X+x_c, y=Y+ \frac{\sqrt{2} \beta\gamma r_c}{2(\sqrt{2}\beta x_c -\gamma)}, z=Z, r=R+r_c$ so that the critical point $E_2$ moves to the origin.  Now we introduce another coordinate system $(x_t,~y_t,~z_t,~r_t)$ in terms of $(X,~Y,~Z,~R)$ so that $J(E_2)$ modifies to its diagonal form.  By using the eigenvectors of $J(E_2)$, we introduce the following coordinate system
\begin{equation}\renewcommand{\arraystretch}{1.5}	
	\begin{bmatrix}
		x_t\\
		y_t\\
		z_t\\
		r_t
	\end{bmatrix}\renewcommand{\arraystretch}{1.5}
	=\begin{bmatrix}
		1 & -\frac{\sqrt{2}(\gamma-\sqrt{2}\beta x_c)^{\frac{3}{2}}}{\beta \sqrt{\gamma}} & 0 &0 \\	
		0 &  1 & 0 & 0\\
		0 & 0 & 1 & 0\\
		0 & 0 & 0 & 1
	\end{bmatrix}\renewcommand{\arraystretch}{1.5}
	\begin{bmatrix}
		X\\
		Y\\
		Z\\
		R
	\end{bmatrix}\label{eqn58}
\end{equation}		
and corresponding to these coordinate system the autonomous system (\ref{auto-Exp-Int-21}) is transformed into	
\begin{equation}	\renewcommand{\arraystretch}{1.5}
	\begin{bmatrix}
		x_t'\\
		y_t'\\
		z_t'\\
		r_t'
	\end{bmatrix}
	=\begin{bmatrix}
		\frac{\sqrt{2}\beta\gamma r_c}{2(\sqrt{2}\beta x_c-\gamma)} & 0 & 0 & 0\\	
		0  &  0 & 0 & 0\\
		0 & 0 & 0 & 0\\
		0&0&0&0
	\end{bmatrix}
	\begin{bmatrix}
		x_t\\
		y_t\\
		z_t\\
		r_t
	\end{bmatrix}		
	+	
	\begin{bmatrix}
		non\\
		lin\\
		ear\\
		terms
	\end{bmatrix}.\label{eqn59}	
\end{equation}	
By center manifold theory there exists a continuously differentiable function 	$h$:$\mathbb{R}^3$$\rightarrow$$\mathbb{R}$ such that
\begin{align}
	x_t=h(y_t,z_t,r_t)=a_1y_t^2+a_2z_t^2+a_3r_t^2+a_4 y_tz_t+a_5y_tr_r+a_6z_tr_t+higher~order~terms\label{eqn60}
\end{align}
Differentiating both side of (\ref{eqn60}) with respect to $N$, we get
\begin{eqnarray}\renewcommand{\arraystretch}{1.5}	
	\frac{dx_t}{dN}=\begin{bmatrix}
		2a_1y_t+a_4z_t+a_5r_t & 2a_2z_t+a_4y_t+a_6r_t & 2a_3r_t+a_5y_t+a_6z_t
	\end{bmatrix}\begin{bmatrix}
		\frac{dy_t}{dN}\\
		\frac{dz_t}{dN}\\
		\frac{dr_t}{dN}
	\end{bmatrix}\label{eqn61}
\end{eqnarray}	
By inserting the expression of $y_t'$, $z_t'$ and $r_t'$ from (\ref{eqn59}) and then also by inserting (\ref{eqn61}) and comparing the coefficient of lowest order terms, we have $a_1=-\frac{3}{\sqrt{2}\beta\gamma}(\gamma-\sqrt{2}\beta x_c)^2, a_2=\frac{x_c^2}{\sqrt{2}\beta\gamma}(\gamma+\sqrt{2}(\alpha-2\beta)x_c)(\sqrt{2}\beta x_c-\gamma),a_3=a_4=a_6=0, a_5=\frac{4x_c}{r_c\sqrt{\gamma}}\sqrt{\gamma-\sqrt{2}\beta x_c}$.  Hence, the center manifold (up to second order) can be written as
\begin{eqnarray}
	x_t=a_1y_t^2+a_2 z_t^2+a_5 r_t y_t +higher~order~terms\label{eqn62}
\end{eqnarray}
and the flow on the center manifold (up to second order) is obtained by
\begin{align}
	y_t'&=\left\{(1-\gamma)(1-r_c)\frac{3\sqrt{\gamma}x_c}{2(\gamma-\sqrt{2}\beta x_c)^\frac{3}{2}}-\frac{\beta \sqrt{\gamma} r_c x_c^2}{\sqrt{2}\sqrt{\gamma-\sqrt{2}\beta x_c}}\right\}z_t^2+higher~order~terms,\label{eqn63}\\
	z_t'&=\left\{\frac{\alpha r_c x_c^2}{\sqrt{2}}+\frac{3\beta x_c^2}{\sqrt{2}(\gamma-\sqrt{2}\beta x_c)}(r_c-1)\right\}z_t^3+higher~order~terms,\label{eqn64}\\
	r_t'&=-\frac{3r_cx_c^2\beta}{\sqrt{2}}(1-r_c)^2z_t^2+higher~order~terms.\label{eqn65}
\end{align}
As the r.h.s. of the expression of the center manifold depends on three coordinates $y_t$, $z_t$ and $r_t$, so here we can only determine the dynamics on the center manifold near the origin by using the flow equations (\ref{eqn63}), (\ref{eqn64}) and (\ref{eqn65}).  For a meaningful choices of $x_c,\beta,\gamma,r_c,\alpha$, the dynamics in the $3$D coordinate system near the origin is shown as in figure \ref{space_of_critical_point}.

\subsubsection{$~Set~of~critical~points~F_2$}

The Jacobian matrix at the critical point $F_2$ corresponding to the autonomous system (\ref{auto-Exp-Int-21}) can be put as
\begin{equation}\renewcommand{\arraystretch}{1.5}	
	J(F_2)=\begin{bmatrix}
		\frac{\sqrt{2}\beta\gamma r_c}{2(\sqrt{2}\beta x_c-\gamma)} & -r_c\sqrt{\gamma}\sqrt{\gamma-\sqrt{2}\beta x_c} & 0&0\\	
		0  &  0 & 0&0\\
		0 & 0 & 0&0\\
		0&0&0&0 
	\end{bmatrix}.\label{eqn66}	
\end{equation}	
The eigenvalues of $J(F_2)$ are $\frac{\sqrt{2}\beta\gamma r_c}{2(\sqrt{2}\beta x_c-\gamma)}$, $0$, $0$ and $0$.  $[1,0,0,0]^T$ be the eigenvector corresponding to the eigenvalue $\frac{\sqrt{2}\beta\gamma r_c}{2(\sqrt{2}\beta x_c-\gamma)}$ and $\left[-\frac{\sqrt{2}(\gamma-\sqrt{2}\beta x_c)^{\frac{3}{2}}}{\beta \sqrt{\gamma}},1,0,0\right]^T$, $[0,0,1,0]^T$ and $[0,0,0,1]^T$ are the eigenvectors corresponding to the eigenvalue $0$.  For a fixed $x_c$ and $r_c$, first we take the shifting transformation $x=X+x_c, y=Y- \frac{\sqrt{2} \beta\gamma r_c}{2(\sqrt{2}\beta x_c -\gamma)}, z=Z, r=R+r_c$ so that the critical point $E_2$ moves to the origin.  Now we introduce another coordinate system $(x_t,~y_t,~z_t,~r_t)$ in terms of $(X,~Y,~Z,~R)$ so that $J(F_2)$ converts to its diagonal form.  By using the eigenvectors of $J(F_2)$, we introduce the following coordinate system
\begin{equation}\renewcommand{\arraystretch}{1.5}	
	\begin{bmatrix}
		x_t\\
		y_t\\
		z_t\\
		r_t
	\end{bmatrix}\renewcommand{\arraystretch}{1.5}
	=\begin{bmatrix}
		1 & \frac{\sqrt{2}(\gamma-\sqrt{2}\beta x_c)^{\frac{3}{2}}}{\beta \sqrt{\gamma}} & 0 &0 \\	
		0 &  1 & 0 & 0\\
		0 & 0 & 1 & 0\\
		0 & 0 & 0 & 1
	\end{bmatrix}\renewcommand{\arraystretch}{1.5}
	\begin{bmatrix}
		X\\
		Y\\
		Z\\
		R
	\end{bmatrix}\label{eqn67}
\end{equation}		
and corresponding to these coordinate system the autonomous system (\ref{auto-Exp-Int-21}) is transformed into	
\begin{equation}	\renewcommand{\arraystretch}{1.5}
	\begin{bmatrix}
		x_t'\\
		y_t'\\
		z_t'\\
		r_t'
	\end{bmatrix}
	=\begin{bmatrix}
		\frac{\sqrt{2}\beta\gamma r_c}{2(\sqrt{2}\beta x_c-\gamma)} & 0 & 0 & 0\\	
		0  &  0 & 0 & 0\\
		0 & 0 & 0 & 0\\
		0&0&0&0
	\end{bmatrix}
	\begin{bmatrix}
		x_t\\
		y_t\\
		z_t\\
		r_t
	\end{bmatrix}		
	+	
	\begin{bmatrix}
		non\\
		lin\\
		ear\\
		terms
	\end{bmatrix}.\label{eqn68}	
\end{equation}	
By center manifold theory there exists a continuously differentiable function 	$h$:$\mathbb{R}^3$$\rightarrow$$\mathbb{R}$ such that
\begin{align}
	x_t=h(y_t,z_t,r_t)=a_1y_t^2+a_2z_t^2+a_3r_t^2+a_4 y_tz_t+a_5y_tr_r+a_6z_tr_t+higher~order~terms\label{eqn69}
\end{align}
Differentiating both side of (\ref{eqn69}) with respect to $N$, we get
\begin{eqnarray}\renewcommand{\arraystretch}{1.5}	
	\frac{dx_t}{dN}=\begin{bmatrix}
		2a_1y_t+a_4z_t+a_5r_t & 2a_2z_t+a_4y_t+a_6r_t & 2a_3r_t+a_5y_t+a_6z_t
	\end{bmatrix}\begin{bmatrix}
		\frac{dy_t}{dN}\\
		\frac{dz_t}{dN}\\
		\frac{dr_t}{dN}
	\end{bmatrix}\label{eqn70}
\end{eqnarray}	
By inserting the expression of $y_t'$, $z_t'$ and $r_t'$ from (\ref{eqn67}) and then also by inserting (\ref{eqn69}) and comparing the coefficient of lowest order terms, we have $a_1=\frac{3}{\sqrt{2}\beta\gamma}(\gamma-\sqrt{2}\beta x_c)^2, a_2=\frac{x_c^2}{\sqrt{2}\beta\gamma}(\gamma+\sqrt{2}(\alpha-2\beta)x_c)(\sqrt{2}\beta x_c-\gamma),a_3=a_4=a_5=a_6=0$.  Hence, the center manifold (up to second order) can be written as
\begin{eqnarray}
	x_t=a_1y_t^2+a_2 z_t^2 +higher~order~terms\label{eqn71}
\end{eqnarray}
and the flow on the center manifold (up to second order) is obtained by
\begin{align}
	y_t'&=-\left\{(1-\gamma)(1-r_c)\frac{3\sqrt{\gamma}x_c}{2(\gamma-\sqrt{2}\beta x_c)^\frac{3}{2}}-\frac{\beta \sqrt{\gamma} r_c x_c^2}{\sqrt{2}\sqrt{\gamma-\sqrt{2}\beta x_c}}\right\}z_t^2+higher~order~terms,\label{eqn72}\\
	z_t'&=\left\{\frac{\alpha r_c x_c^2}{\sqrt{2}}+\frac{3\beta x_c^2}{\sqrt{2}(\gamma-\sqrt{2}\beta x_c)}(r_c-1)\right\}z_t^3+higher~order~terms,\label{eqn73}\\
	r_t'&=-\frac{3r_cx_c^2\beta}{\sqrt{2}}(1-r_c)^2z_t^2+higher~order~terms.\label{eqn74}
\end{align}

\begin{figure}[h]
	\centering
	\includegraphics[width=0.8\textwidth]{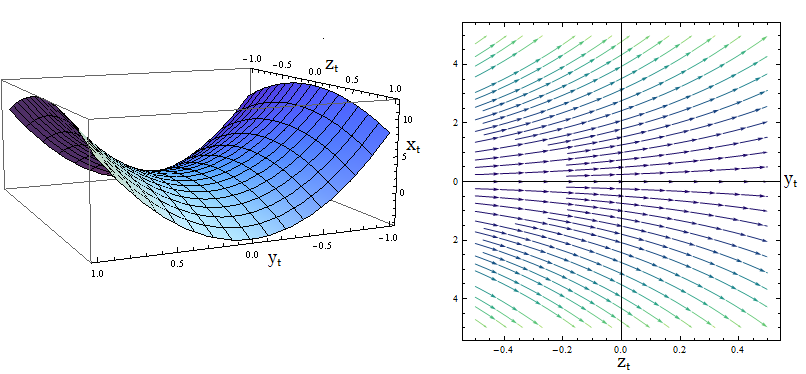}
	\caption{Left hand side figure is the plot of the center manifold corresponding to the critical point $F_2$ and the projection of the vector field in arbitrary small neighbourhood of the origin on $y_tz_t$ plane is shown as at right hand side.   For existence the domain of critical point and to satisfy the condition (\ref{eqn18}) we have drawn this $3$-dimensional phase plot for $x_c=-1,\beta=1,\gamma=2,r_c=2,\alpha=1$.  From the plot of the vector field we conclude that the vector field is unstable about the $y_t$ axis.}
	\label{F_21}
\end{figure}

\begin{figure}[h]
	\centering
	\includegraphics[width=0.8\textwidth]{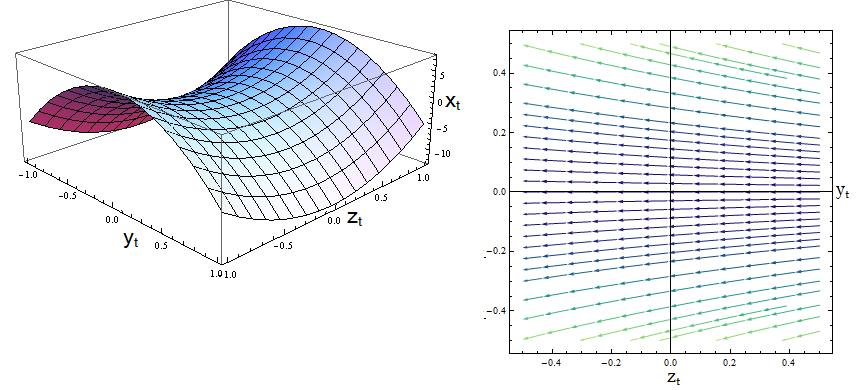}
	\caption{Left hand side figure is the plot of the center manifold corresponding to the critical point $F_2$ and the projection of the vector field in arbitrary small neighbourhood of the origin on $y_tz_t$ plane is shown as at right hand side.  For existence the domain of critical point and to satisfy the condition (\ref{eqn18}) we have drawn the center manifold and the phase plot for $x_c=1,\beta=-1,\gamma=2,r_c=2,\alpha=1$.  From the plot of the vector field we conclude that for this case also the vector field is unstable about $y_t$ axis.}
	\label{F_22}
\end{figure}

The plot of the center manifolds for $r_t=0$, that means., in $x_ty_tz_t$ coordinate system are shown as in figure \ref{F_21} and figure \ref{F_22} for several choices of $x_c,\beta,\gamma,r_c$ and $\alpha$.  The projection of the vector field in arbitrary small neighbourhood of the origin on $y_tz_t$ plane corresponding to the same choices of $x_c,\beta,\gamma,r_c$ and $\alpha$ are also shown as in figure \ref{F_21} and figure \ref{F_22}.

\newpage
%
%




\section{Compactification and the dynamics around the points at infinity} \label{infinity}
The idea to compactify the space $\mathbb{R}^n$ by addition of points at infinity and to map them into finite points is frequently used in the setting of two-dimensional differential equations \cite{inf1,inf2,inf2:a}.  An early study of analyzing the global behavior of a planar dynamical system via compactification was carried out by Bendixson using the stereographic projection of the sphere (often called Bendixson sphere) onto the plane.  However, this stereographic projection has the drawback of obliterate the``different directions at infinity'' \cite{inf3}.  French mathematician Henri Poincar\'{e} overcame this difficulty by projecting $\mathbb{R}^2$ onto the Poincar\'{e} hemisphere through its center.

In this compactification scheme we draw straight lines starting from the center of the unit sphere $S^2=\{(X,Y,Z)\in \mathbb{R}^3\mid X^2+Y^2+Y^2=1\}$ to $xy$-plane which is tangent to $S^2$ at either the north or south pole.    It is to be noted that each straight line meets the unit sphere once and the $xy$-plane once.  Let a straight line intersects the sphere at $(X', Y', Z')$ and the $xy$-plane at $(x',y')$, then our  projective transformation is realized by the transformation
\begin{align}
	x'=\frac{X'}{Z'},~y'=\frac{Y'}{Z'} \label{p1}
\end{align}

So an arbitrary point $(x,y)$ on the $xy$-plane can be expressed uniquely by a point $(X,Y,Z)$ on upper/lower half of the sphere as (\ref{p1}) \cite{inf3,inf4}.  This scheme has the advantage that the critical points of different directions at infinity are spread out along the equator of the sphere.  

Given an autonomous system of differential equations on $\mathbb{R}^2$ 
\begin{align}
	&\dot{x}=P(x,y) \label{p2}	\\ &  \dot{y}=Q(x,y) \label{p3}
\end{align}
where $P$ and $Q$ are polynomial functions of $x$ and $y$.   We can write (\ref{p2}) and (\ref{p3}) in the form of a single differential equation 
$$
\frac{dy}{dx}=\frac{Q(x,y)}{P(x,y)}
$$
which yields
\begin{align}
	Q(x,y)dx-P(x,y)dy=0\label{p4}
\end{align}

Now for any point $(x,y)$ using (\ref{p1}) we obtain
\begin{align}
	dx=\frac{ZdX-XdZ}{Z^2},~dy=\frac{ZdY-YdZ}{Z^2}\label{p5}
\end{align}
Hence, the differential equation (\ref{p4}) can be expressed as
\begin{align}
	Q(ZdX-XdZ)-P(ZdY-YdZ)=0 \label{p5a}
\end{align}	
where
$$
P=P(x,y)=P\left(X/Z,Y/Z\right)
$$	
and	
$$
Q=Q(x,y)=Q\left(X/Z,Y/Z\right).
$$

Let $m$ be the maximum degree of the terms in $P$ and $Q$.  We multiply both side of the equation (\ref{p5a}) by $Z^m$  to eliminate $Z$ from the denominator and obtain	
\begin{align}
	ZQ^*dX-ZP^*dY+(YP^*-XQ^*)dZ=0\label{p6}
\end{align}
where
$$
P^*(X,Y,Z)=Z^mP(X/Z,Y/Z)
$$	
and
$$
Q^*(X,Y,Z)=Z^mQ(X/Z,Y/Z)
$$	
are polynomials in $(X,Y,Z)$.  

The equator of $S^2$ can be expressed by $\{(X,Y,0 ) | X^2+Y^2=1\}$.  So, the critical points of (\ref{p6}) on the equator of $S^2$ where $Z = 0$ are given by the equation	
\begin{align}
	XQ^*-YP^*=0\label{p7}
\end{align}
On the equator of the Poincar\'{e} sphere we can derive
\begin{eqnarray}
	XQ^*-YP^* &= & XQ_m(X,Y)-YP_m(X,Y) \\
	&=& \mathcal{S}~~ \mbox{(say)} \nonumber,
\end{eqnarray}
where $P_m$ and $Q_m$ are homogeneous $j^{th}$ degree polynomials in $x$ and $y$of the system $(\ref{p2})-(\ref{p3})$.  So the critical points at infinity are the set 
$\{(X, Y, 0)\mid X^2 + Y^2 = 1 ~\mbox{and}~ \mathcal{S}=0 \}$.  

For the autonomous system (23), we obtain the equation (\ref{p7}) as 
\begin{equation}
	XQ_9(X,Y)-YP_9(X,Y)=\frac{3}{2}X^5Y^5 =  0 \label{p8a}
\end{equation}
So there are critical points on the equator of $S^2$ at $\pm \left(1,0,0 \right)$ and at $\pm \left(0,1,0 \right)$.  One may note that the flow on the equator of $S^2$ is clockwise for $XY<0$ and counter-clockwise for $XY>0$.  We also note that the flow on $S^2$ defined by (\ref{p6}) is topologically equivalent at the antipodal points of $S^2$ as $m~ (=9)$ is odd \cite{perko}.

\begin{figure}[h!]
	\centering
	\includegraphics[width=0.4\textwidth]{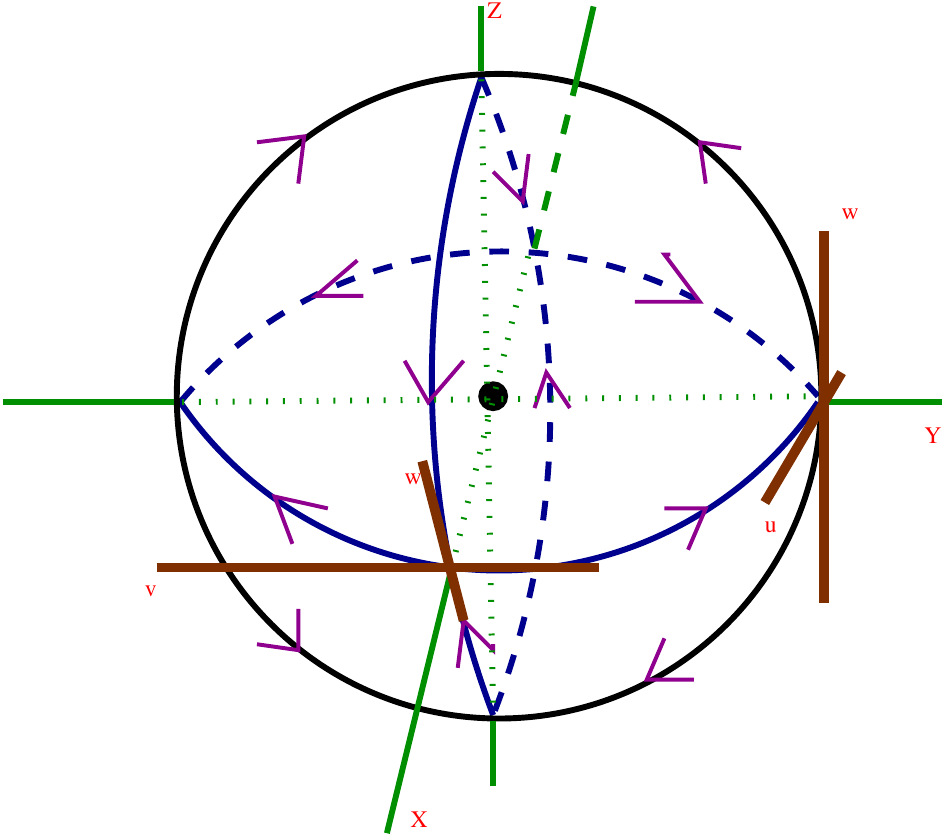}\label{PS3}
	\caption{Flow on Poincar\'{e} sphere.}
	\label{sphere}
\end{figure}

Now to analyze the vector field near the critical point $\left(1,0,0 \right)$, we project the neighborhood $X>0$  onto $vw$-plane ($X=1$, see figure \ref{sphere}) which is tangent to the Poincar\'{e} sphere $S^2$ at $\left(1,0,0 \right)$.  The flow near $\left(1,0,0 \right)$ is topologically equivalent to the flow near the origin of the following system

\begin{align}
	\begin{split}
		\dot{v}&=vw^mP\left(\frac{1}{w},\frac{v}{w}\right)-w^m Q\left(\frac{1}{w},\frac{v}{w}\right)	\\ 
		\dot{w}&=w^{m+1}P\left(\frac{1}{w},\frac{v}{w}\right)
		\label{poin1}
	\end{split}
\end{align}
where,\\
\begin{eqnarray*}
	P \left(\frac{1}{w},\frac{v}{w}\right) &=& 3\frac{v^2}{w^5}+\sqrt{\frac{3}{2}}s\left(\frac{v^3}{w^5}+\frac{v^3}{w^7}\right)+3\eta \left(\frac{1}{w}-\frac{v^2}{w^3}-\frac{v^2}{w^5} \right),\\
	Q\left(\frac{1}{w},\frac{v}{w}\right) &=& -\sqrt{\frac{3}{2}}s\frac{v^4}{w^7}-\frac{3}{2}\left(\frac{v^3}{w^5}-\frac{v^5}{w^7}+\frac{v^5}{w^9} \right)
\end{eqnarray*}

\begin{figure}
	\centering
	\subfigure[]{%
		\includegraphics[width=7cm,height=6cm]{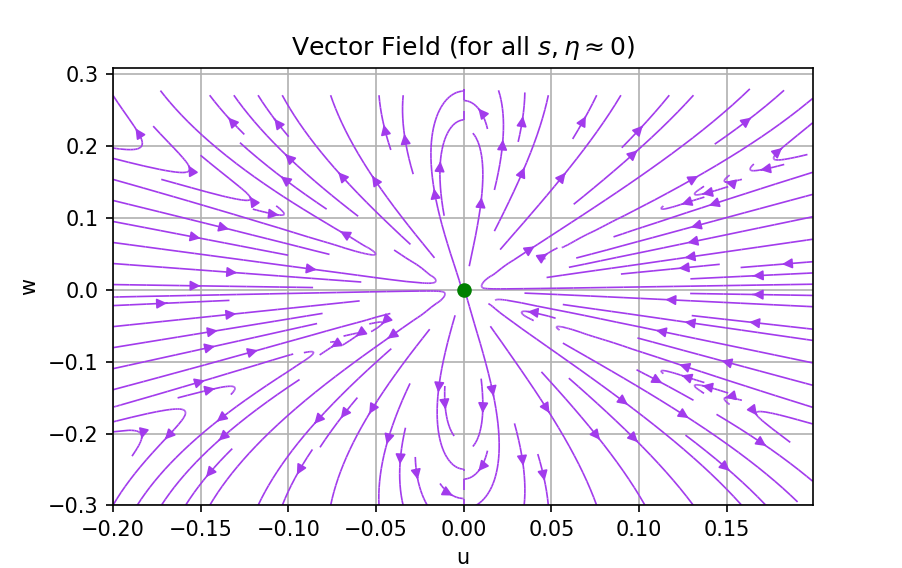}\label{PS1}}
	\qquad
	\subfigure[]{%
		\includegraphics[width=7cm,height=6cm]{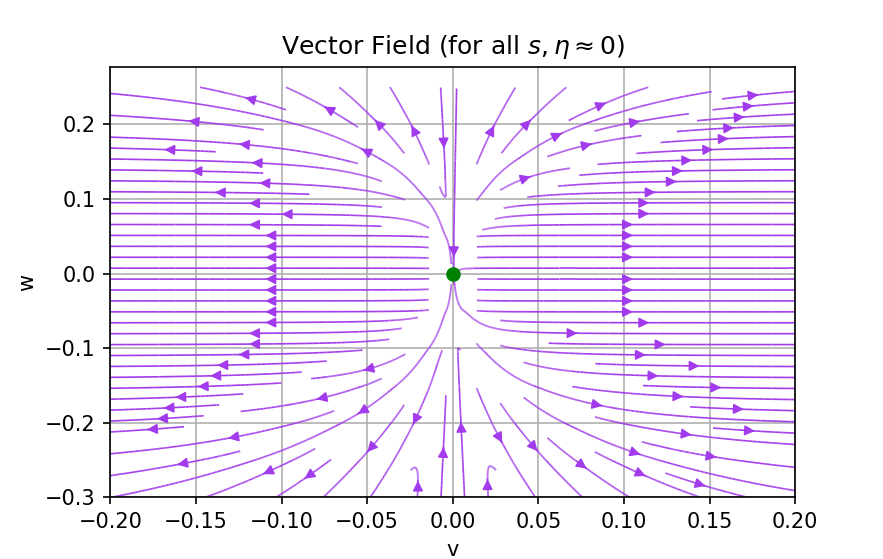}\label{PS2}}
	\caption{Vector fields are shown near $(0,\pm 1,0)$ in figure \ref{PS1} and near $(\pm 1,0,0)$  in figure \ref{PS2}.  All the critical points at infinity are saddle in nature. }
	\label{PS1vPS2}
\end{figure}

Similarly, we project the  neighborhood $Y>0$ onto $uw$-plane ($Y=1$, see figure \ref{sphere}) which is tangent to the Poincar\'{e} sphere $S^2$ at $\left(0,1,0 \right)$.  The flow near $\left(0,1,0 \right)$ is topologically equivalent to the flow near the origin of the following system
\begin{align}
	\begin{split} 
		\dot{u}&=uw^mQ\left(\frac{u}{w},\frac{1}{w}\right)-w^mP\left(\frac{u}{w},\frac{1}{w}\right)\\
		\dot{w}&=w^{m+1}Q\left(\frac{u}{w},\frac{1}{w}\right)
		\label{poin2}
	\end{split}
\end{align}
where,\\
\begin{eqnarray*}
	P \left(\frac{u}{w},\frac{1}{w}\right) &=& 3\frac{u^3}{w^5} +\sqrt{\frac{3}{2}}s\left(\frac{u^2}{w^5}+\frac{u^4}{w^7}\right)+3\eta \left(\frac{u}{w}-\frac{u}{w^3}-\frac{u^3}{w^5} \right),\\
	Q\left(\frac{u}{w},\frac{1}{w}\right) &=& -\sqrt{\frac{3}{2}}s\frac{u^3}{w^7}-\frac{3}{2}\left(\frac{u^2}{w^5}-\frac{u^2}{w^7}+\frac{u^4}{w^9} \right)
\end{eqnarray*}
It is to be noted that origin is a nonhyperbolic critical point for both systems (\ref{poin1}) and (\ref{poin2}).  The jacobian matrix at the origin contains two zero eigenvalues for both systems, i.e., origin is multiple critical point.  So center manifold theory is not applicable and we instead resort to the method of numerical plotting the vector field (see figures \ref{PS1} and \ref{PS2}).   We find that both critical points $\left(1,0,0 \right)$ and $\left(0,1,0 \right)$ are saddle in nature and so their antipodal points (see figure \ref{sphere} for the flow on the Poincar\'{e} sphere).   

To analyze points at infinity of an autonomous system with three variables we have to consider $S^3 ~\in ~\mathbb{R}^4$.   Similarly, for $4D$ autonomous system we need to consider $S^4~\in ~ \mathbb{R}^5$.  A complete survey of Poincar\'{e} sphere in $4D$ or $5D$ space, which entails elaborate mathematical illustrations, is beyond the scope of this paper (see \cite{perko} for an overview).

\section{Cosmological implications}\label{Cosmological Implications}

\subsection{Interaction model 1}

The set of critical points $A_1$ represents the universe dominated by both the dark matter (DM) and the scalar field (DE) where scalar field behaves like a cosmological constant fluid ($\omega_{\phi}=-1$). This set is a normally hyperbolic set and stability this set can be completely classified by considering the sign of the non-zero eigenvalues in the corresponding eigen-directions. Also, this statement is verified by center manifold theory as we have shown in eq.(\ref{flow A1}) in section \ref{CMT}. The stability depends on the flow parallel to the  eigen-direction of $\mu_1$ (i.e. x-axis). The DM dominated universe is realized on the set for $y_c\longrightarrow 0$ and DE dominated universe for $y_c\longrightarrow 1$ (since the effective equation of state $\omega_{eff}=-y^{2}_c$, where $0\leq y^{2}_c\leq 1$).
A dust dominated universe ($\omega_{eff}=0$, $q=\frac{1}{2}$) is obtained for $y_c=0$ and may be future attractor as for $\eta>0$.
Therefore, the set can predict dust dominated future decelerated universe for $\eta>0$ and the dust dominated past attractor for $\eta<0$. 
However, the universe dominated by the scalar field is represented by the points on the set $A_1$ are evolving in quintessence regime at late times for $\eta>0$. It agrees with the observation ($-0.56<q<-0.49$) for $0.812<y_c<0.841$ and $\eta>0$ (see Table \ref{Physical parameters model1}). Figure \ref{A1stable} for $\eta=0.1$ and $s=1$ shows that the magenta colored line $A_1$ is stable attractor where as for $\eta=-0.1$ and $s=1$ the figure \ref{A1unstable} shows that the set $A_1$ is unstable in the phase plane. \\

Scalar field dominated solutions $B_1$ and $C_1$ correspond to the late time stable attractors in quintessence era with $\Omega_{\phi}=1$, so coincidence problem cannot be alleviated by these points. The evolution of late time accelerated solutions $B_1$ and $C_1$ are attracted in quintessence era either for 
($\eta \leq \frac{1}{6}~~\mbox{and}~~ \left(-\sqrt{2}<s<0~~\mbox{or}~~ 0<s<\sqrt{2}\right)$) or for ($\frac{1}{6}<\eta <\frac{1}{2}~~\mbox{and}~~ \left(-\sqrt{3-6 \eta }<s<0~~\mbox{or}~~ 0<s<\sqrt{3-6 \eta }\right)$).  These equilibrium points are conceded by observations ($-0.56<q<-0.49$) for the parameter bounds $ \bigg[\eta \leq 0.33 $ and $ (-1.01<s<-0.938 $ or $ 0.938<s<1.01)\bigg] $ or $ \bigg[0.33<\eta <0.353$ and $ (-\sqrt{3-6\eta}<s<-0.938$ or $0.938<s<\sqrt{3-6\eta })\bigg]  $. Therefore, the late time accelerated evolutions of the universe attracted in quintessence era only. Figure \ref{A1stable}, for $\eta=0.1, s=1$ and the figure \ref{A1unstable}  for $\eta=-0.1, s=1$ confirm that the points  $B_1$ and $C_1$ are accelerated late time stable attractors in quintessence era. Figure  \ref{Evolution_C} refers to the evolution of cosmological parameters for $\eta=0.01$ and $s=-1$ showing that the evolution of the accelerated universe is attracted at late times in the quintessence era.  \\

Critical points $D_1$ and $E_1$ represent DM-DE scaling solutions with $0<\Omega_{\phi}<1$ in the phase plane therefore, the points can alleviate the coincidence problem. The points are hyperbolic type, so linear stability theory is sufficient to describe the nature of the points. For $\eta=\frac{1}{2}$, the points $D_1$ and $E_1$ describe the de Sitter expansions of the universe ($\Omega_{\phi}=1, \Omega_{m}=0, \omega_{eff}=-1$) but both the points are unstable there. For $\eta=0$, the points can predict the future decelerated dust dominated universe ( the physical parameters are: $\omega_{eff}=0$, $q=\frac{1}{2}$) and for this case, DE behaves as dust (since $\omega_{\phi}=0$) and the points are stable for $s^2>3$. That is for an uncoupled model ($\eta=0$), we have obtained a future DE (like dust) dominated decelerated universe in dust era. Figure \ref{FstableDeceleration} shows that the dust dominated decelerated scaling attractor is represented by the points  $D_1$ and $E_1$ for $\eta=0, s=-1.89$. It is worth mentioning that the points $D_1$ and $E_1$ show physically interesting nature when coupling of interaction constrained as $\frac{1}{6}<\eta<\frac{1}{2}$. Despite this, depending on other parameter restrictions, both the points correspond to late time accelerated scaling attractors with ratio of energy densities of DE and DM being unity which can alleviate the coincidence problem. Figure (\ref{D1stable}, \ref{E1stable}) shows for the parameters  $\eta=0.2, s=-1.89$ that  the points represent accelerated scaling attractor with  $\Omega_{\phi}\approx0.7, \Omega_{m}\approx0.3, \omega_{eff}\approx-0.4$ and $q\approx-0.1$ which agrees the present observed accelerated universe. More precisely, according to the present observations ($-0.56<q<-0.49$) the restriction for the accelerated universe in parameter space is $ \bigg[ 0.33<\eta <0.353$ and $(-\frac{\sqrt{3}}{2} \sqrt{\frac{4 \eta-12 \eta ^2+1}{\eta }}<s<-\sqrt{3-6\eta}$ or $\sqrt{3-6\eta}<s<\frac{\sqrt{3}}{2} \sqrt{\frac{4 \eta-12 \eta ^2+1}{\eta }})\bigg] $. Also, figure \ref{Evolution_E} with same parameter values confirms that the late time accelerated evolution of the universe is attracted in quintessence era.
For both the cases, late time accelerated attractor is achieved satisfying $\frac{\Omega_{\phi}}{\Omega_{m}}\approx O(1)$ which can alleviate the coincidence problem successfully.

\subsection{Interaction model 2}

For the interaction model 2, we have obtained several number of non-isolated set of critical points in 4D phase-space. DE behaves as cosmological constant for all the sets. Although the accelerated de Sitter like solutions namely, the sets $B_2$ and $C_2$ (with $\Omega_{\phi}=1, \Omega_{m}=0, \omega_{eff}=-1, q=-1$) represent the evolution of the universe dominated by potential of the scalar field. Due to having no non-vanishing eigenvalues, the center manifold theory is not applicable for the stability analysis of $B_2$ and $C_2$. In the section \ref{CMT}, it has been shown  that the sets $B_2$ and $C_2$ are special cases of the set $A_2$ and  their stability are similar to that of $A_2$ (for $y_c=\pm 1$) which has been shown in the figure \ref{vector field A_2}. In conclusion, the sets of points $B_2$ and $C_2$ exhibit the unstable de Sitter like solutions in phase space.  The
matter-scalar field scaling solutions are described by the sets of critical points namely $E_2$ and $F_2$ which are non-hyperbolic in nature and by center manifold theory, the sets also represent the transient era in the evolution of the universe (see figures \ref{space_of_critical_point},\ref{F_21} and \ref{F_22}). Thus, the interaction model 2 produces non-hyperbolic sets of critical points which are either dominated by scalar field or representing the scaling solutions. All the solutions are saddle in nature and exhibit the transient nature of evolution of the universe.

\section{Summary and Concluding remarks}\label{conclusions}

We have studied the cosmological dynamics of interacting scalar field in the background of spatially flat FLRW metric where the scalar field plays a role of DE candidate and pressureless dust as DM. Two interaction models have been considered here. In the first model, potential $V(\phi)$ of scalar field as well as the coupling term $\lambda(\phi)$ of scalar field are taken as inverse square form. Whereas in the second interaction model, the potential and coupling term are taken as the exponential function of scalar field $\phi$. Since the cosmological evolution equations are complicated in nature, we have performed dynamical systems analysis to achieve the qualitative behaviour of the cosmological model. We have obtained hyperbolic type critical points from model 1 and non-hyperbolic sets from model 2. Linear stability theory is sufficient to provide the nature of hyperbolic critical points and center manifold theory is employed  to obtain exact dynamical nature of the points on the sets. Classical stability is also investigated by finding the adiabatic sound speed for the models and corresponding conditions for classical stability are shown in table \ref{classical stability modelQ1} and the table \ref{classical stability modelQ2} for the interacting model 1 and interacting model 2 respectively. \\

In the interaction model 1, where $Q\propto H \rho_{m}$, we choose potential $V(\phi)$ as well as coupling function $\lambda(\phi)$ of the scalar field to be 
$V(\phi)=\lambda(\phi)=V_0 \phi^{-2}$, where $V_0$ is a constant. As a result of which, the dynamical variables $y$ and $z$ are not independent at all. Eventually the system (\ref{Gen_autonomous}) reduces to a 2D autonomous system (\ref{autonomous2D}) (since $y=z$) and consequently $s=u$. We have procured a normally hyperbolic set of critical points $A_1$ in this autonomous system and
in cosmologically viable region, the vector field near $A_1$ bifurcates at $\eta=0$ (see in figures \ref{A1stable} and \ref{A1unstable}). In section \ref{CMT}, we have also employed Center manifold theory to manifest that the stability of the normally hyperbolic set ($A_1$) depends on the eigen direction of non-vanishing eigenvalue.
However, from numerical computation we observed that points on the set $A_1$ has interesting nature in cosmological point of view. The points on the set representing the scalar field dominated universe at late times when $\eta>0$ and $y_c\neq 0$ and representing the dust dominated decelerated unstable universe for $\eta<0$ ($y_c= 0$). It should be noted that the set $A_1$ represents the dust dominated future decelerated universe for $\eta>0$ and $y_c=0$.\\

Hyperbolic critical points such as $B_1$ and $C_1$ are identical in all cosmological perspective. The points are scalar field dominated stable solutions in the phase plane. It is worthy to note that late time accelerated universe is attracted in quintessence era described by the points $B_1$ and $C_1$ (see figures \ref{A1stable} and \ref{A1unstable}). \\

Critical points $D_1$ and $E_1$ are also hyperbolic in nature and we have employed the linear stability theory to characterise the vector field. Although, the points describe the accelerated de Sitter solution for $\eta=\frac{1}{2}$, but they are unstable there ($\Omega_{m}=0, \Omega_{\phi}=1, \omega_{eff}=q=-1$). Further, the points delineate late time (stable) decelerating dust dominated ($\omega_{eff}=0, q=\frac{1}{2}$) universe for $\eta=0$ and for this case, DE behaves as dust ($\omega_{\phi}=0$). Accordingly, the system can predict the future decelerated model of the universe for the uncoupled case. On the other hand, the points $D_1$ and $E_1$ are physically interesting at late times only for $\eta<\frac{1}{2}$. For some restrictions on parameters, the points represent the late time accelerated scaling attractors satisfying the order of energy densities of DE and DM to be unity and alleviating the coincidence problem (see figures \ref{D1stable} and \ref{E1stable}).
It is interesting to note that the figures \ref{Evolution_C} and \ref{Evolution_E} of time evolution of cosmological parameters $\Omega_{m}, \Omega_{\phi}, \omega_{eff}, q$ suggest that the late time evolution of the universe is achieved in quintessence era for this interaction scenario. \\

In the interacting model 2, the interaction term $Q\propto \rho_{m}$ has been chosen locally  (free from Hubble parameter $H$) and the potential of scalar field and the coupling function of scalar field are considered as exponential type in such a way that the dynamical variables $y$ and $z$ are independent. As a result, the system  (\ref{Gen_autonomous}) immerses in 4D autonomous system (\ref{auto-Exp-Int-21}) of real space where the Hubble parameter $H$ is considered to be an additional dynamical variable. In this interacting scenario, we have obtained a collection of non-isolated set of critical points, all of which are non-hyperbolic in nature but not normally hyperbolic sets. So, the Center Manifold Theorem is studied in detail (in the section \ref{CMT}) to realize the cosmological dynamics of evolution of the sets. We have obtained de Sitter like solutions dominated by potential of scalar field having transient nature of evolution. At the same time, we have also acquired matter-scalar field scaling solutions having the transient nature in the phase space.\\

In conclusions, one may summarize that the phase space analysis of the model 1 with inverse square potential reveals the late time accelerated attractor solutions where the dark energy and dark matter density parameters are of same order (satisfying $0<\Omega_{m}<1$ and $0<\Omega_{\phi}<1$) which leads to alleviation of the cosmic coincidence problem. On the other hand, interaction model 2 with exponential potential can provide only saddle like solutions in phase space which cannot give the possible solution of the coincidence problem.


\section*{Acknowledgments}
The author G. Mandal is supported by UGC, Govt. of
India through Senior Research Fellowship [Award Letter No. F.82-
1/2018(SA-III)] for Ph.D. S. Chakraborty is grateful to CSIR, Govt. of India for giving Senior Research Fellowship (CSIR Award No: 09/096(1009)/2020-EMR-I) for the Ph.D work.  We would like to express our deep gratitude to Professor Subenoy Chakraborty for his enthusiastic encouragement and useful critiques to write this manuscript. We thank the annonymous Reviewers for their careful reading of our manuscript and their many insightful comments.


\end{document}